%% file: ms.tex
\documentclass[12pt]{article}

\usepackage{booktabs} % For formal tables

\usepackage[ruled]{algorithm2e} % For algorithms

\SetAlFnt{\small}
\SetAlCapFnt{\small}
\SetAlCapNameFnt{\small}
\SetAlCapHSkip{0pt}
\IncMargin{-\parindent}

\usepackage{url}

\usepackage{listings}
\usepackage{lstlinebgrd}
\usepackage{color}

\usepackage{multirow}

\usepackage{amssymb}

\usepackage{capt-of}

\definecolor{codegreen}{rgb}{0,0.6,0}
\definecolor{codegray}{rgb}{0.5,0.5,0.5}
\definecolor{codepurple}{rgb}{0.58,0,0.82}
\definecolor{backcolour}{rgb}{0.95,0.95,0.92}
\definecolor{lightblue}{rgb}{0.67,0.84,0.90}

\lstdefinestyle{mystyle}{
	backgroundcolor=\color{backcolour},
	commentstyle=\color{codegreen},
	keywordstyle=\color{magenta},
	numberstyle=\tiny\color{codegray},
	stringstyle=\color{codepurple},
	basicstyle=\scriptsize,
	breakatwhitespace=false,
	breaklines=true,
	captionpos=t,
	keepspaces=true,
	numbers=left,
	numbersep=5pt,
	showspaces=false,
	showstringspaces=false,
	showtabs=false,
	tabsize=2,
	mathescape=true
}

\usepackage{authblk}
\usepackage{todonotes}

\usepackage{graphics}

\usepackage{graphicx}
\usepackage{caption}
\usepackage{subcaption}

\usepackage{pifont}
\newcommand{\cmark}{\ding{52}}%
\newcommand{\xmark}{\ding{56}}%

\usepackage{tikz}
\newcommand*\circled[1]{\tikz[baseline=(char.base)]{
		\node[shape=circle,draw,inner sep=0.5pt] (char) {#1};}}

\begin{document}
% Title portion. Note the short title for running heads
%\title[Synergy: A Transparent Consolidated HW/SW Framework for CNNs on Embedded FPGAs]{Synergy: A Transparent Consolidated HW/SW Framework for High Throughput CNNs on Embedded FPGAs}
%\title[Synergy: A Transparent Consolidated HW/SW Framework for CNNs on Embedded FPGAs]{Synergy: A Transparent Consolidated HW/SW Framework for High Throughput CNNs on Embedded Heterogeneous SoC}

%\title[Synergy: A HW/SW Framework for CNNs on Embedded Heterogeneous SoC]{Synergy: A HW/SW Framework for High Throughput CNNs on Embedded Heterogeneous SoC}

\title{\vspace*{-9ex}Synergy: A HW/SW Framework for High Throughput CNNs on Embedded Heterogeneous SoC}
\author{Guanwen Zhong\thanks{zhguanwen@gmail.com}}
\author{Akshat Dubey\thanks{akshatdubey@nus.edu.sg}}
\author{Tan Cheng\thanks{tancheng@comp.nus.edu.sg}}
\author{Tulika Mitra\thanks{tulika@comp.nus.edu.sg}}
\affil{School of Computing, National University of Singapore}

%\author[$\dagger$]{Guanwen Zhong}
%\author[$\star$]{Akshat Dubey}
%\author[$\ddag$]{Tan Cheng}
%\author[$\diamond$]{Tulika Mitra}
\date{\vspace{-6ex}}

\maketitle

\begin{abstract}
Convolutional Neural Networks (CNN) have been widely deployed in diverse application domains. There has been significant progress in accelerating both their training and inference using high-performance GPUs, FPGAs, and custom ASICs for datacenter-scale environments. The recent proliferation of mobile and IoT devices have necessitated real-time, energy-efficient deep neural network inference on embedded-class, resource-constrained platforms. In this context, we present {\em Synergy}, an automated, hardware-software co-designed, pipelined, high-throughput CNN inference framework on embedded heterogeneous system-on-chip (SoC) architectures (Xilinx Zynq). {\em Synergy} leverages, through multi-threading, all the available on-chip resources, which includes the dual-core ARM processor along with the FPGA and the NEON SIMD engines as accelerators. Moreover, {\em Synergy} provides a unified abstraction of the heterogeneous accelerators (FPGA and NEON) and can adapt to different network configurations at runtime without changing the underlying hardware accelerator architecture by balancing workload across accelerators through work-stealing. {\em Synergy} achieves 7.3X speedup, averaged across seven CNN models, over a well-optimized software-only solution. {\em Synergy} demonstrates substantially better throughput and energy-efficiency compared to the contemporary CNN implementations on the same SoC architecture.
\end{abstract}

%\begin{CCSXML}
%	<ccs2012>
%	<concept>
%	<concept_id>10010520.10010553</concept_id>
%	<concept_desc>Computer systems organization~Embedded and cyber-physical systems</concept_desc>
%	<concept_significance>500</concept_significance>
%	</concept>
%	<concept>
%<concept_id>10010520.10010521.10010542.10010294</concept_id>
%<concept_desc>Computer systems organization~Neural networks</concept_desc>
%<concept_significance>500</concept_significance>
%</concept>
%<concept>
%<concept_id>10010520.10010521.10010542.10010543</concept_id>
%<concept_desc>Computer systems organization~Reconfigurable computing</concept_desc>
%<concept_significance>500</concept_significance>
%</concept>
%	</ccs2012>
%\end{CCSXML}
%
%\ccsdesc[500]{Computer systems organization~Embedded and cyber-physical systems}
%\ccsdesc[500]{Computer systems organization~Neural networks}
%\ccsdesc[500]{Computer systems organization~Reconfigurable computing}


%\keywords{Hardware/software co-design, CNNs, FPGAs, heterogeneous computing, accelerator abstraction, multi-threading, work stealing}

%\renewcommand{\shortauthors}{G. Zhong et al.}

%\input{samplebody-journals}

\input{intro.tex}

\input{background.tex}
\input{relatedwork.tex}

\input{framework.tex}

\input{experiment.tex}

\input{conclusion.tex}

%\begin{acks}
%	reserved. aaaaa
%	%The authors would like to thank Dr. Maura Turolla of Telecom
%	%Italia for providing specifications about the application scenario.
%	%
%	%The work is supported by the \grantsponsor{GS501100001809}{National
%	%  Natural Science Foundation of
%	%  China}{http://dx.doi.org/10.13039/501100001809} under Grant
%	%No.:~\grantnum{GS501100001809}{61273304\_a}
%	%and~\grantnum[http://www.nnsf.cn/youngscientists]{GS501100001809}{Young
%	%  Scientists' Support Program}.	
%\end{acks}

% Bibliography
\bibliographystyle{plain}
\bibliography{sample-bibliography}

\end{document}

%% file: intro.tex
\section{introduction}
\label{sec:intro}
%\todoinline{
%1. Heterogeneous architectures\\
%2. CNN intro\\
%3. Parallelism inside CNN\\
%4. work imbalance\\
%5. contribution in this work}
%\todoinline{add google TPU reference}
\begin{sloppypar}

Convolutional Neural Networks (CNNs) are a popular class of deep learning method with wide range of applications, including computer vision, image/video processing, natural language processing, and others. A typical CNN consists of multiple layers. Given an application, such as image classification, the network is first trained with the training dataset. The trained network is then deployed for inference, i.e., classification of a new image. Both the training and the inference are compute- and memory-intensive, but also exhibit massive intrinsic parallelism. Thus, there exist numerous efforts to improve the performance and the energy-efficiency of CNN implementations through architectures and computing substrates that support extensive parallelism, such as GPUs, FPGAs, or even ASIC accelerators. This line of research has primarily focused on the high-performance computing platforms in datacenters or clouds.

%such as convolutional, normalization, pooling, activation, and fully-connected

\begin{figure}[t!]
	\centering
	\includegraphics[width=0.7\columnwidth]{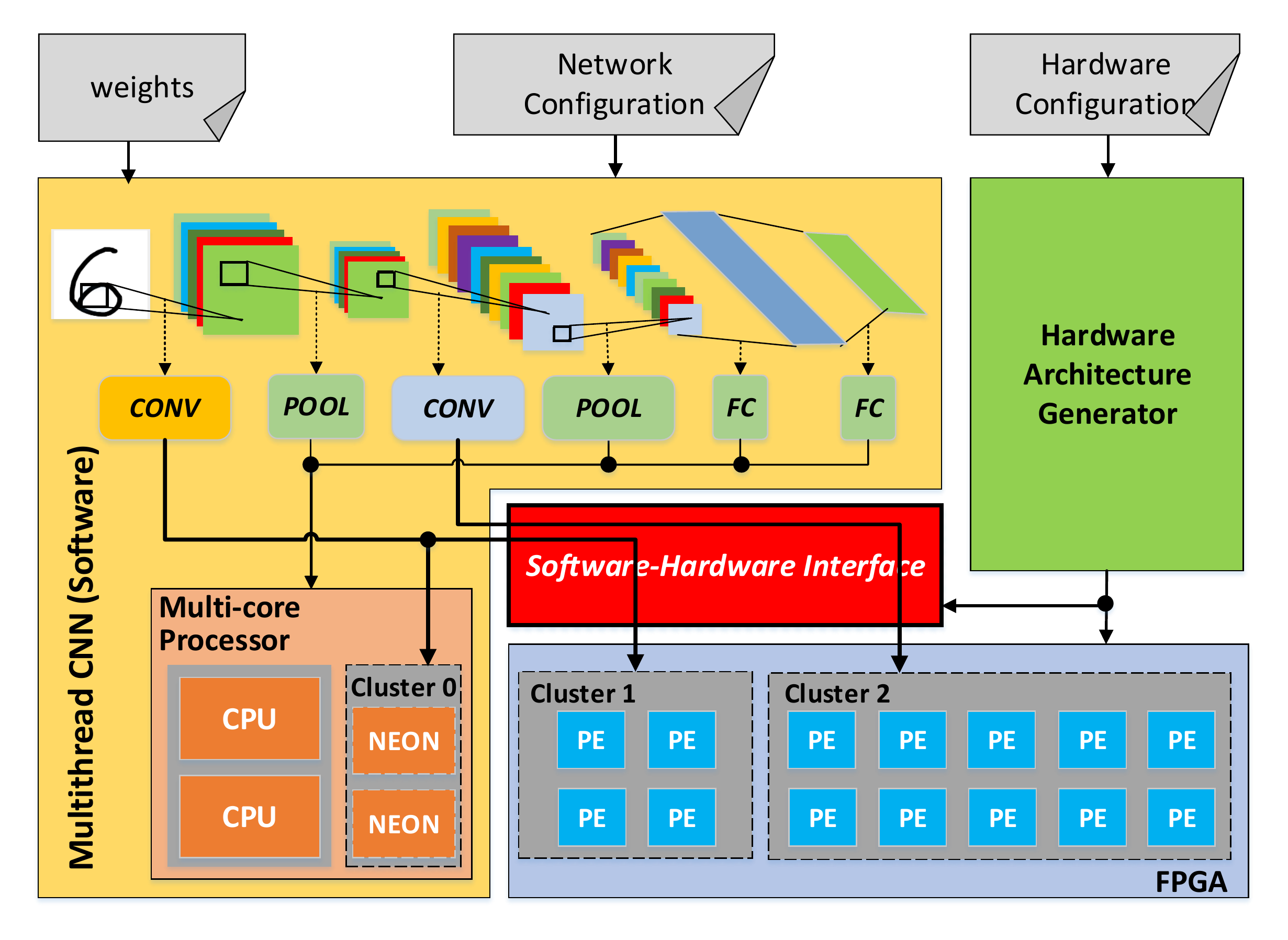}
	\caption{Synergy: Mapping CNNs on Heterogeneous SoC}
	\label{fig:general_mapping}
	%\vspace{-1.5em}
\end{figure}

The proliferation of the mobile devices and the recent emergence of the IoT (Internet of Things) have transformed the computing landscape. There is a compelling need to realise real-time, energy-efficient CNN inference on resource-constrained mobile and IoT edge devices. However, an efficient implementation of CNN-based inference on embedded platforms remains challenging given the resource limitations. In this context, we present {\em Synergy}, an automated, transparent, pipelined, high-throughput, hardware-software co-designed CNN inference framework on embedded heterogeneous SoC architectures. We design {\em Synergy} prototype on the Xilinx Zynq XC7Z020 device leveraging all its available on-chip compute resources, namely the dual-core ARM processor with NEON SIMD engines and the FPGA. {\em Synergy} is a complete system-level solution including a multi-threaded software component, multi-threaded FPGA and NEON accelerators, an interface between hardware and software components, support for dynamic workload balancing, as well as an architecture generator for customized solutions (if required). Figure~\ref{fig:general_mapping} depicts the {\em Synergy} framework mapping a CNN model on a heterogeneous SoC. {\em Synergy} distinguishes itself from the state-of-the-art along multiple dimensions.

{\bf Heterogeneous HW/SW Acceleration}: {\em Synergy} leverages all the compute resources available on a heterogeneous SoC for maximum performance. The convolutional (referred to as CONV hereafter) layers are the most compute-intensive component of the CNN consuming more than 90\% of the execution time~\cite{Zhang_FPGA15_cnn}. All contemporary CNN designs~\cite{caffepresso_cnn}\cite{fpgaConvNet}\cite{DeepBurning} on the Xilinx Zynq platform offload the CONV layers to the FPGA. We observe that the NEON SIMD (Single-Instruction Multiple-Data) engines in the ARM cores are quite effective in accelerating the CONV layers as well. Therefore, harnessing the compute power of the FPGA in conjunction with the NEON engines can reduce the execution latency of the CONV layers significantly. Embracing the heterogeneity --- {\em hardware accelerators on FPGA and software accelerators on NEON} --- for a single computational kernel is a difficult proposition. {\em Synergy} effectively transforms the computation of a CONV layer into a tiled matrix multiplication operation, where the tiles can be processed independently. {\em Synergy} then seamlessly feeds the tiles to both the hardware and the software accelerators to exploit the parallelism within CONV layers. {\em Synergy} improves the overall latency and throughput by 12\% and 15\% respectively, averaged across multiple CNN models, using NEON and FPGA compared to FPGA-only solutions.

{\bf Transparent Consolidation of Accelerators}: Most contemporary FPGA-based CNN frameworks~\cite{FP-DNN}\cite{caffepresso_cnn}\cite{going_deeper_cnn}
\cite{Suda_OpenCL_cnn}\cite{fpgaConvNet}\cite{DeepBurning}\cite{Zhang_FPGA15_cnn}\cite{Caffeine_iccad2016} rely heavily on customizing the CONV layer accelerators for each individual network to minimize latency. The configuration of a CNN (number and type of layers, nodes, connections, etc.) is dependent on the application. Given a specific CNN model, existing approaches perform an automated (or manual) design space exploration (DSE) to identify the optimal accelerator architectures for the CONV layers of that network on the target FPGA device. This approach has the drawback that the application developer needs to be involved in the DSE and High-Level Synthesis (HLS) to map the given CNN model on the FPGA. Even if the DSE and HLS steps can be fully automated, there are still some quirks~\cite{Shen_RP_CNN_ISCA} that make this process quite challenging for an application developer with limited experience in FPGAs. Second, in an embedded FPGA device with strict resource constraints, a single accelerator design is used by all the CONV layers of the network in a time-multiplexed fashion even though the different layers have diverse compute requirements. In other words, the single accelerator is a compromise to offer the best average performance across all the CONV layers of the network, but it is not ideal for any particular CONV layer~\cite{Shen_RP_CNN_ISCA}. Moreover, this single CONV accelerator is still custom-designed for each network through extensive DSE and HLS.

In contrast, {\em Synergy} accelerators (FPGA, NEON) are network-agnostic. A fixed set of accelerators is used irrespective of the network and layer as the CONV layer computation is transformed into tiled matrix multiplications. Using fine-grained tiled matrix multiplication operations as fundamental primitives (as opposed to complete CONV layer) in conjunction with a {\it work-stealing} software scheduler that distributes these tiles to the different accelerators and balances the workload, {\em Synergy} achieves comparable performance to the customized network-specific implementations returned through DSE. Thus, {\em Synergy} can bypass the DSE and HLS for each individual CNN model and provide a completely out-of-the-box solution.

%An exception to this approach is~\cite{angel_eye}, where a fixed hardware architecture is synthesized on the FPGA, but a compiler statically generates a set of instructions (describing the process of CNN execution) that execute on the fixed architecture.

{\bf High-Throughput HW/SW Multi-Threaded Pipeline:} The transparent consolidation of heterogeneous HW/SW accelerators in {\em Synergy} provides a powerful abstraction layer for any CNN implementation. The abundance of sensors on mobile and IoT devices capturing continuous data streams demand in-situ real-time inference (e.g., continuous object detection in image/video stream~\cite{streaming_dnn}). Here throughput (i.e., frames per second) is the defining metric as opposed to minimizing the latency for each individual frame in isolation. {\em Synergy} employs a HW/SW multi-threaded pipelined design of the different layers of the network that allows consecutive frames (images) from the streaming input to be processed concurrently exploiting inter-frame parallelism and improving throughput. However, in this pipelined design, CONV layers from different frames may need to be processed simultaneously on the FPGA. This inter-frame parallelism is easy to support in {\em Synergy} as the different matrix multiplications from the different CONV layers simply generate matrix multiplication tiles and the tiles from different layers get distributed in a transparent fashion to the set of heterogeneous accelerators to be executed in parallel. {\em Synergy} achieves 39.5 -- 136.4 frames/second throughput and consumes 14.4 -- 55.8 mJ/frame energy depending on the CNN model. This is substantially better than the contemporary CNN implementations on the Xilinx Zynq XC7Z020 device (see Table~\ref{tab:energy_comp}). Moreover, the concurrent execution of multiple CONV layers on FPGA accelerators in {\em Synergy} greatly improves their utilization. The low utilization of the accelerators is a critical issue in full-system implementation of CNNs on resource-constrained devices where non-compute intensive layers (e.g., pooling, activation and fully-connected) implemented in software on not so powerful CPUs (e.g., ARM) take up significant execution time while the FPGA remains inactive. The pipelined design with multiple frames in-flight keeps the accelerators busy in {\em Synergy} with 99.8\% average utilization.

{\bf Automated Customized Accelerator Design}: {\em Synergy} offers a default plug-n-play solution for a na{\"i}ve application developer that avoids the complex DSE and HLS for each individual CNN model. An experienced designer, on the other hand, may want to further improve the performance by designing accelerators that are optimized for the CNN model at hand. {\em Synergy} offers an automated approach to customized acceleration design for a specific CNN model. The framework accepts the network configuration (number and type of layers) corresponding to a CNN model as an input. The designer only needs to provide the architectural parameters for the matrix multiplication accelerators. The {\em Synergy} framework not only can automatically synthesize the accelerators (according to designer-defined parameters) on the FPGA fabric but also generate the appropriate hardware-software interface to synergistically engage these newly synthesized accelerators. This is shown as the ``Hardware Architecture Generator'' in Figure~\ref{fig:general_mapping}. In addition, the same architecture generator can be used to synthesize the accelerators and the HW/SW interface for a new SoC device. Thus, {\em Synergy} provides a complete push-button solution to CNN acceleration on a heterogeneous platform.

\end{sloppypar}

%% file: relatedwork.tex
\section{related works}
%\todoinline{Need to update related works}
\begin{sloppypar}
	
The state-of-the-art FPGA-based CNN works are shown in Table~\ref{tab:related_work}. To the best of our knowledge, there is no work focusing on heterogeneous HW/SW acceleration (with CPUs, NEONs and FPGA) for CNNs. We classify the existing works into two categories: {\it network-dependent} and {\it network-independent} FPGA-based CNN frameworks. 

%\vspace{0.3em}
{\bf Network-dependent Frameworks} generally require designers to explore different configurations to generate a hardware architecture for a specific CNN network manually or with the help of scripts provided, and perform synthesis (which normally takes half to one hour) to generate the bitstream. Given a new network, designers need to redo the above steps, which is time consuming. This approach is well-explored and can produce extreme high-performance CNN accelerators, but sacrificing the flexibility to different networks. Almost all the existing FPGA-based CNN works~\cite{FP-DNN}\cite{caffepresso_cnn}\cite{MT_CNN}\cite{going_deeper_cnn}\cite{Shen_RP_CNN_ISCA}\cite{Suda_OpenCL_cnn}\cite{FINN_FPGA17}\cite{fpgaConvNet_fccm16}\cite{fpgaConvNet}\cite{DeepBurning}\cite{Caffeine_iccad2016}\cite{cnn_multi_fpga}\cite{Zhang_FPGA15_cnn} use the {\it network-sensitive} approach. \cite{Shen_RP_CNN_ISCA}\cite{Suda_OpenCL_cnn}\cite{cnn_multi_fpga}\cite{Zhang_FPGA15_cnn} require designers to manually explore architectures for different networks, while \cite{FP-DNN}\cite{FINN_FPGA17}\cite{fpgaConvNet_fccm16}\cite{fpgaConvNet}\cite{DeepBurning}\cite{Caffeine_iccad2016} propose automated toolchains for mapping CNNs on FPGAs. \cite{FP-DNN}\cite{caffepresso_cnn}\cite{going_deeper_cnn}\cite{Suda_OpenCL_cnn}\cite{fpgaConvNet}\cite{DeepBurning}\cite{Caffeine_iccad2016}\cite{Zhang_FPGA15_cnn} mainly focus on exploiting {\em intra-frame parallelism} within layers and execute layers in a CNN in sequence, ignoring {\em inter-frame parallelism} across layers. \cite{FINN_FPGA17}\cite{cnn_multi_fpga} map all layers onto FPGAs and enable hardware pipelining to exploit the {\em inter-frame parallelism}. \cite{FINN_FPGA17} proposed an automated framework to accelerate binarized CNNs. Their work maps all layers in an binarized CNN on FPGA and enables hardware pipelining. \cite{cnn_multi_fpga} proposes an approach by mapping convolutional, normalization, pooling, activation and fully-connected layers onto multiple FPGA devices (1 Xilinx ZC706 + 6 Xilinx VC709 boards). Each device is in charge of a specific one or more CNN layers and devices are connected in a ring network. The pipelining flow is controlled by dual-core ARM processors on the Xilinx Zynq board. However, the cost of the deeply pipelined FPGA cluster is too high as it requires multiple high-end FPGA devices and the setup is difficult. Different from \cite{FINN_FPGA17}\cite{cnn_multi_fpga}, \cite{MT_CNN} starts with multi-threaded CNN inference codes and converts all its layers into FPGA accelerators. However, the workload of different layers in a CNN could be imbalanced, which leads to low accelerator utilization, wasting the precious FPGA resources. \cite{Shen_RP_CNN_ISCA} statically splits single large processing engine (PE) used to accelerate convolutional layers into multiple small PEs. Their approach can allow multiple layers running simultaneously with different image frames. However, the evaluation of their work is based on simulation and the performance (execution cycles) of PEs is obtained by Vivado HLS. The performance number is not accurate as it does not consider the runtime overhead of the real platform. 	
	
\begin{table}[t]
	\centering
	\caption{Current State-of-the-art vs. Synergy}
	\label{tab:related_work}
	\resizebox{0.9\columnwidth}{!}{
		\begin{tabular}{lcccccc}
			\hline
			\multicolumn{1}{|l|}{Reference}              & \multicolumn{1}{c|}{Automated} & \multicolumn{1}{c|}{\begin{tabular}[c]{@{}c@{}}Inter-\\ frame\end{tabular}} & \multicolumn{1}{c|}{\begin{tabular}[c]{@{}c@{}}Self-\\balancing\end{tabular}} & \multicolumn{1}{c|}{\begin{tabular}[c]{@{}c@{}}HW\\ Reuse*\end{tabular}} & \multicolumn{1}{c|}{\begin{tabular}[c]{@{}c@{}}Network \\ Agnostic\end{tabular}}  & \multicolumn{1}{c|}{\begin{tabular}[c]{@{}c@{}}On-board\\ Evaluation\end{tabular}}\\ \hline
			\multicolumn{1}{|l|}{\cite{Zhang_FPGA15_cnn} {[}FPGA'15{]}}    & \multicolumn{1}{c|}{\xmark}        & \multicolumn{1}{c|}{\xmark}                                                                                                             & \multicolumn{1}{c|}{\xmark}                                                            & \multicolumn{1}{c|}{\xmark}  &  \multicolumn{1}{c|}{\xmark}  & \multicolumn{1}{c|}{\cmark}                                               \\ \hline
			\multicolumn{1}{|l|}{\cite{going_deeper_cnn} {[}FPGA'16{]}}      & \multicolumn{1}{c|}{\xmark}        & \multicolumn{1}{c|}{\xmark}                                                     & \multicolumn{1}{c|}{\xmark}                                                            & \multicolumn{1}{c|}{\cmark}  & \multicolumn{1}{c|}{\xmark}  & \multicolumn{1}{c|}{\cmark}                                              \\ \hline						
			\multicolumn{1}{|l|}{\cite{Suda_OpenCL_cnn} {[}FPGA'16{]}}      & \multicolumn{1}{c|}{\xmark}        & \multicolumn{1}{c|}{\xmark}                                                     &  \multicolumn{1}{c|}{\xmark}                                                            & \multicolumn{1}{c|}{\xmark}  & \multicolumn{1}{c|}{\xmark}  & \multicolumn{1}{c|}{\cmark}                                              \\ \hline		
			\multicolumn{1}{|l|}{\cite{caffepresso_cnn} {[}CASES'16{]}}      & \multicolumn{1}{c|}{\xmark}       & \multicolumn{1}{c|}{\xmark}                                                     & \multicolumn{1}{c|}{\xmark}                                                            & \multicolumn{1}{c|}{\cmark}  & \multicolumn{1}{c|}{\xmark}  & \multicolumn{1}{c|}{\cmark}                                               \\ \hline
			\multicolumn{1}{|l|}{\cite{DeepBurning} {[}DAC'16{]}}      & \multicolumn{1}{c|}{\cmark}       & \multicolumn{1}{c|}{\xmark}                                                     & \multicolumn{1}{c|}{\xmark}                                                            & \multicolumn{1}{c|}{\xmark}   & \multicolumn{1}{c|}{\xmark}  & \multicolumn{1}{c|}{\cmark}                                              \\ \hline
			\multicolumn{1}{|l|}{\cite{Caffeine_iccad2016} {[}ICCAD'16{]}}   & \multicolumn{1}{c|}{\cmark}       & \multicolumn{1}{c|}{\xmark}                                                     & \multicolumn{1}{c|}{\xmark}                                                            & \multicolumn{1}{c|}{\xmark}     & \multicolumn{1}{c|}{\xmark}  & \multicolumn{1}{c|}{\cmark}                                            \\ \hline
			\multicolumn{1}{|l|}{\begin{tabular}[l]{@{}l@{}}\cite{fpgaConvNet_fccm16}{[}FCCM'16{]}\\ \cite{fpgaConvNet}{[}FPGA'17{]}\end{tabular}} & \multicolumn{1}{c|}{\cmark}       & \multicolumn{1}{c|}{\xmark}                                                     &  \multicolumn{1}{c|}{\xmark}                                                            & \multicolumn{1}{c|}{\xmark}   & \multicolumn{1}{c|}{\xmark}  & \multicolumn{1}{c|}{\cmark}                                             \\ \hline
			\multicolumn{1}{|l|}{\cite{FINN_FPGA17} {[}FPGA'17{]}}     & \multicolumn{1}{c|}{\cmark}       & \multicolumn{1}{c|}{\cmark}                                                     & \multicolumn{1}{c|}{\xmark}                                                            & \multicolumn{1}{c|}{\xmark}  & \multicolumn{1}{c|}{\xmark}  & \multicolumn{1}{c|}{\cmark}                                              \\ \hline		
			\multicolumn{1}{|l|}{\cite{FP-DNN} {[}FCCM'17{]}}     & \multicolumn{1}{c|}{\cmark}       & \multicolumn{1}{c|}{\xmark}                                                     & \multicolumn{1}{c|}{\xmark}                                                            & \multicolumn{1}{c|}{\cmark}  & \multicolumn{1}{c|}{\xmark}  & \multicolumn{1}{c|}{\cmark}                                              \\ \hline
			\multicolumn{1}{|l|}{\cite{cnn_multi_fpga} {[}ISLPED'16{]}}  & \multicolumn{1}{c|}{\xmark}        & \multicolumn{1}{c|}{\cmark}                                                    &  \multicolumn{1}{c|}{\xmark}                                                            & \multicolumn{1}{c|}{\xmark}  & \multicolumn{1}{c|}{\xmark}  & \multicolumn{1}{c|}{\cmark}                                               \\ \hline
			\multicolumn{1}{|l|}{\cite{MT_CNN} {[}SOCC'17{]}}      & \multicolumn{1}{c|}{\xmark}       & \multicolumn{1}{c|}{\cmark}                                                    & \multicolumn{1}{c|}{\xmark}                                                            & \multicolumn{1}{c|}{\xmark}  & \multicolumn{1}{c|}{\xmark}  & \multicolumn{1}{c|}{\cmark}                                               \\ \hline
			\multicolumn{1}{|l|}{\cite{Shen_RP_CNN_ISCA} {[}ISCA'17{]}}     & \multicolumn{1}{c|}{\xmark}         & \multicolumn{1}{c|}{\cmark}                                                    & \multicolumn{1}{c|}{\xmark}                                                            & \multicolumn{1}{c|}{\cmark}  & \multicolumn{1}{c|}{\xmark}  & \multicolumn{1}{c|}{\xmark}                                                \\ \hline
			\multicolumn{1}{|l|}{\cite{angel_eye} {[}TCAD'17{]}}      & \multicolumn{1}{c|}{\cmark}        & \multicolumn{1}{c|}{\xmark}                                                     & \multicolumn{1}{c|}{\xmark}                                                            & \multicolumn{1}{c|}{\cmark}  & \multicolumn{1}{c|}{\cmark}  & \multicolumn{1}{c|}{\cmark}                                              \\ \hline \hline
			\multicolumn{1}{|c|}{\begin{tabular}[c]{@{}c@{}}\bf Proposed\\\bf Synergy\end{tabular}}       & \multicolumn{1}{c|}{\cmark}       & \multicolumn{1}{c|}{\cmark}                                                    & \multicolumn{1}{c|}{\cmark}                                                           & \multicolumn{1}{c|}{\cmark}   & \multicolumn{1}{c|}{\cmark}  & \multicolumn{1}{c|}{\cmark}                                             \\ \hline
			\multicolumn{7}{l}{HW Reuse: different CONV layers and FC layers can share the same FPGA accelerators}                                                                                                                                                                                                                                                                                                                                                                         
			%\vspace{-2.5em}
		\end{tabular}
	}
\end{table}

{\bf Network-independent Frameworks} leverage a fixed optimized hardware architecture for various CNN networks. To adapt to different networks and achieve good hardware efficiency, this approach relies on either static (compiler) or dynamic (runtime scheduler) techniques. The key advantage of this approach is that designers can easily switch different networks at runtime without going through the time-consuming synthesis step. \cite{angel_eye} belongs to this category. Their approach relies on a compiler developed to statically generate a set of instructions (describing the process of CNN execution) that execute on the fixed hardware architecture. Layers are executed in sequence in their work. Moreover, as \cite{angel_eye} includes data quantization to reduce memory requirement, their approach can support large networks on embedded FPGA devices. However, their approach can not allow multiple layers running concurrently with different input frames, which might result in low accelerator utilization.

\vspace{0.3em}
Synergy supports {\it network-independent} feature. More specifically, we propose a hardware abstraction to unify various computing elements (NEON cores and FPGA) within an FPGA-based SoC architecture. Thus, Synergy can leverage all computing elements (multiple ARM cores, its NEON cores and FPGA) to accelerate CNNs via HW/SW multi-threading, unleashing the true power of heterogeneity. Different from~\cite{angel_eye}, Synergy adapts to various networks by leveraging a {\it work-stealing} scheduler (Section~\ref{subsec:stealing}) in software to automatically balance the workload of accelerators at runtime without changing hardware or software implementations. Moreover, Synergy provides an automated toolchain to allow designers to explore various accelerator architectures or migrate designs to other embedded FPGA devices.

\vspace{-1em}
%Different from existing approaches, we modify the CNN framework, Darknet, with multi-threaded support and propose a hardware/software multi-threading approach, Synergy, to improve throughput of CNNs by fully leveraging all computing elements (multiple ARM cores, its NEON cores and FPGA) within single FPGA-based SoC architecture. To improve FPGA utilization and solve workload imbalance problems among layers of a CNN, we introduce a dynamic workload balancing technique, work stealing (mentioned in Section~\ref{subsec:stealing}), in our hardware/software co-design. The whole framework including software (multi-threaded CNNs, scheduling library for accelerators, host code and Linux kernels) and hardware (HLS integration, interfaces and memory subsystem) generation is automated and evaluated on real platforms (Xilinx ZC702 and Zedboard) running on Linux.

%Different from existing approaches, we propose a hardware/software multi-threading approach to exploit {\em inter-layer parallelism} within single FPGA-based SoC platform. Moreover, due to imbalance workload among convolutional layers in a CNN model, we develop a work stealing scheduler to fully utilize our hardware accelerators.
\end{sloppypar}

%% file: framework.tex
\section{The Synergy Framework}
\label{sec:synergy_framework}
%\todoinline{1. Add one general framework figure including both HW/SW flows;\\2. briefly introduce how the system works for CNNs}
%\todoinline{We need to mention {\bf Generality} of Synergy in somewhere}

%The Synergy framework, as shown in Figure~\ref{synergy_framework}, comprises mainly of the multi-threaded CNN framework, the various HW accelerators, and the HW/SW multi-threading library.
%\todoinline{Make sure we include FC layer in acceleration later if we have time, otherwise, donot say we accelerate FC layer}
%\todoinline{Remember to put Section Number in the below text!!}
%\todoinline{Change HW ACC to PE!!}

%\todoinline{Need to put one paragraph to bring the general picture of Synergy}
\begin{sloppypar}
{\em Synergy}, as shown in Figure~\ref{fig:general_mapping}, is an automated framework to map CNN models onto embedded FPGA-based heterogeneous SoC platforms. {\em Synergy} targets the CNN inference phase and successfully unleashes the power of heterogeneity of the SoC architectures by leveraging all its compute elements (CPUs, NEON engines, and FPGA).

A CNN model contains multiple layers such as convolutional, normalization, pooling, activation and fully connected layers. The input of a layer is the output of the previous layer. When input frames stream into the CNN, the layers can process different frames concurrently. This {\it inter-frame parallelism} can be exploited to improve throughput.

{\em Synergy} uses the FPGA logic and the NEON engines to accelerate the most compute-intensive layers (CONV) in a CNN, while the CPU cores work on the other layers (such as pooling, activation and fully-connected layers) and preprocessing functions (e.g., normalization, scaling and data layout transformation). As shown in Figure~\ref{fig:general_mapping}, a designer can instantiate multiple processing engines (referred to as PE hereafter) on the FPGA to accelerate the CONV layers. The computation in a CONV layer is transformed into a set of independent tiled matrix-multiplication operations, called jobs as mentioned in Section~\ref{subsubsec:conv_layer}. These jobs are executed by the FPGA and the NEON accelerators in parallel.

To improve the inference throughput and accelerator utilization, {\em Synergy} supports HW/SW multi-threaded pipeline where the CPU cores and the accelerators work on different layers of different frames concurrently. Therefore, we modify the traditional single-threaded CNN framework with multi-threaded support. Specifically, the workload in each layer is conducted by the corresponding thread and the communication between layers is performed through a mailbox (a synchronized first-in-first-out buffer) accessible by the threads. Multiple threads collaborate with each other in a producer-consumer fashion constructing the full network.

As multi-threading is a software concept, hardware accelerators cannot directly share the well-established mechanisms in the multi-threading model such as mutex, semaphore, and mailbox. To abstract away the hardware accelerators as hardware threads and extend operating system to support HW/SW threads, we adapt ReconOS~\cite{ReconOS}, an open-source operating system for reconfigurable computing. {\em ReconOS} provides the HW/SW multi-threading technique that we build upon to accelerate CNN applications on FPGA-based heterogeneous SoC platforms. Each hardware accelerator or PE is represented by a {\em delegate thread} in the software space that behaves just like traditional software threads as shown in Figure~\ref{fig:sw_flow} and explained in detail in Section~\ref{subsubsec:delegate_thread}.

The accelerators (PEs and NEONs) can be grouped into multiple clusters so that each CONV layer can have its own private cluster. However, {\em Synergy} accelerators are not customized given a specific CNN model. Thus the generic multi-cluster configuration may not be optimal for each network and may lead to imbalance in execution time of the different CONV layers. {\em Synergy} employs work-stealing (detailed in Section~\ref{subsec:stealing}), a dynamic workload scheduling technique, to deal with the workload imbalance among the clusters. The jobs (independent tiled matrix-multiplication operations) from the different CONV layers are distributed to the different accelerator clusters. An idle cluster steals workload from the other busy clusters and thereby achieves workload balance across clusters and maximizes throughput. Within a cluster, the jobs are dispatched to the available accelerators (NEONs and FPGA-based PEs) in a round-robin fashion.

%Synergy leverages NEON cores in ARM processor and FPGA-based accelerators to accelerate CONV layers and uses CPUs for other layers and preprocessing functions. For a CONV layer, its convolutional operations can be represented by matrix multiplication (MM). As input size of MM is large, Synergy splits the MM into tiles. We represent workload of each tile calculation as a job. Each CONV thread has a courier function to deliver jobs to a {\it job queue}. An accelerator cluster (explained in Section~\ref{subsubsec:conv_layer} takes jobs from the job queue and dispatch to its available accelerators (NEONs and FPGA-based PEs).

%
%different layers can have their own private accelerator cluster. This spatial partitioning strategy (cluster configuration) has a great impact on the performance of CONV layers. Since workload of the CONV layers is imbalanced, there exists a design space exploration problem to find the proper cluster configuration such that the execution time of each cluster is balanced. This spatial partitioning strategy is a static mapping approach. However, the static mapping approach requires extensive effort to profile the performance of cluster configurations for CONV layers with different input size, which is time-consuming.

The {\em Synergy} framework provides a default architecture of the FPGA-based PEs and their cluster configuration that has been designed to provide quality performance across a range of CNN models. These clusters and their constituent PEs are pre-synthesized in {\em Synergy} corresponding to each unique FPGA device/platform and do not need to be generated for each individual CNN model. In other words, the FPGA bitstream remains unchanged across different CNN models and the FPGA device need not be reconfigured corresponding to each CNN model unless desired by the application developer. Given a CNN model corresponding to an application, {\em Synergy} takes in a network configuration file that defines the architecture of the CNN as input. The CPU-based host code used to control the hardware accelerators, Linux kernels and HW/SW multi-threaded library are written as templates. With the network configuration file and the software templates, {\em Synergy} automatically generates a software/hardware multi-threaded CNN application in C.

If an advanced application developer wants to customize the PE and cluster design for a specific CNN model, the {\em Synergy} framework offers a hardware architecture generator (Section~\ref{subsubsec:generator}). In this case, {\em Synergy} takes in a hardware configuration file as input and creates the hardware architecture by instantiating the HLS-RTL accelerator templates in C corresponding to the tiled matrix-multiplication operations. These FPGA-based accelerators for the CONV layers are generated by a commercial HLS tool from the C templates, while accelerator interfaces and memory subsystem are created by RTL templates. Both the generation of software and hardware components are completely automated.

\end{sloppypar}

\input{softwareflow.tex}

\input{hardwareflow.tex}
\input{selfbalancing.tex}

%% file: softwareflow.tex
\subsection{Software Architecture}
\label{subsec:sw_comp}
\begin{figure}[t!]
	\centering
	\includegraphics[width=0.75\columnwidth]{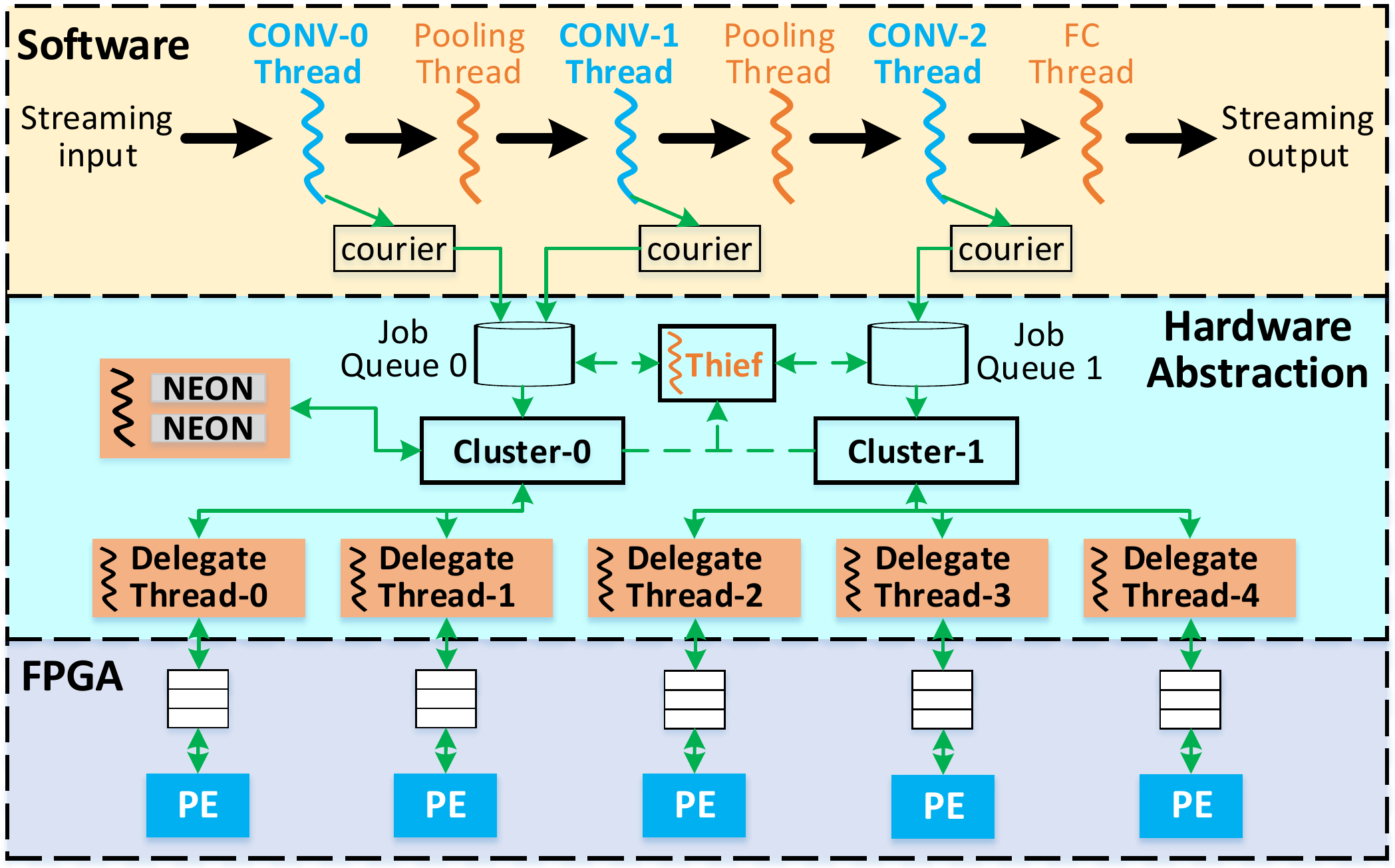}
	\caption{Overview of the Software Architecture}
	%\vspace{-1.5em}
	\label{fig:sw_flow}
\end{figure}

%\todoinline{Remember to add a figure for multi-threading CNN with mailbox}
\begin{sloppypar}
Figure~\ref{fig:sw_flow} shows the software component in {\em Synergy}. We explain the functionality in software to implement the CONV layers and the other layers, preprocessing functions.
%There exist many popular open-source deep neural network frameworks that support CNNs such as Caffe~\cite{caffe} and Darknet~\cite{darknet13}. As Darknet does not depend on any external library and it is fairly easy to install it on embedded platforms, we extend it with multi-threaded support.

%In the following subsections, we explain how to deal with different layers on heterogeneous architectures and how to control hardware accelerators with software threads in detail.

%In the following subsections, we explain how to deal with different layers on heterogeneous architectures, how to control hardware accelerators in software threads and how the overall software flow works in detail.
\end{sloppypar}

\subsubsection{\bf CONV Layers}
\label{subsubsec:conv_layer}
CONV layers are the most compute-intensive components in a CNN, occupying  more than 90\% of the execution time during inference~\cite{Zhang_FPGA15_cnn}. They take in input feature maps and convolve them with convolutional filters to obtain output feature maps. As we target the low-end embedded platforms, the FPGA resources are quite limited and we cannot generate a dedicated accelerator for each convolutional layer in a CNN model like~\cite{going_deeper_cnn}\cite{Zhang_FPGA15_cnn}\cite{Caffeine_iccad2016}\cite{cnn_multi_fpga}. Therefore, in our implementation, we need to share the hardware accelerators among the convolutional layers.
We transform the convolution operations into matrix multiplication (MM) by flattening and rearranging the input features~\cite{FP-DNN}\cite{Suda_OpenCL_cnn}. A data layout transformation is required to convert the 3D array in CONV layers into 2D array, which is known as {\it image-to-column} ({\it im2col}) function in many popular open-source CNN frameworks such as Caffe~\cite{caffe} and Darknet~\cite{darknet13}. Details related to the data layout transformation can be found in~\cite{caffe}\cite{Suda_OpenCL_cnn}. {\em Synergy} leverages both the FPGA-based PEs and the NEON engines to accelerate the MM computations.
%Popular open-source deep neural network frameworks like Caffe~\cite{caffe} and Darknet~\cite{darknet13} convert convolution computation to matrix multiplication (MM) by flattening and rearranging the input features~\cite{conv_as_mm}. For portability and simplicity, we directly map MM in convolutional layers on FPGAs. In this section, we first present our FPAG-based accelerators and discuss several design challenges.

%Then, we introduce the architecture of hardware interface that is used to abstract accelerators and communicate with delegate threads~\cite{ReconOS} in user space. In the last subsection, we describe the automated generator to create hardware support for Synergy.

%\todoinline{Should move to SW part}
\begin{lstlisting}[language=C, caption=Tiled Matrix Multiplication, basicstyle=\tiny, label={lst:tiled_mm}, linebackgroundcolor={\ifnum\value{lstnumber}>4 \ifnum\value{lstnumber}<15\color{lightblue}\fi\fi}, escapechar=|]
/* Tile Size: TS; Loop bounds: N, M, K */
Tile-t1: for(t1=0;t1<floor(N/TS);++t1){
  Tile-t2: for(t2=0;t2<floor(M/TS);++t2){
    ... //Initialization
    tiled_mm: for(t3=0;t3<floor(K/TS);++t3){ |\label{line:tiledMM1}|
      //Step 1: Copy data from DDR to local buffer(a,b,c);
      data_copy(ddr_a, ddr_b, a, b, offsetA, offsetB);
      //Step 2: Kernel Computation
      loop1: for(i=0;i<TS;++i){
        loop2: for(j=0;j<TS;++j){
          loop3: for(k=0;k<TS;++k){
            c[i][j] += a[i][k]*b[k][j]; }}}}
     //Step 3: Write data from local buffer to DDR
    data_send(c,ddr_c,offsetC); |\label{line:tiledMM2}|
}}
\end{lstlisting}

After flattening and rearranging the input features, the input matrices of the matrix multiplication in convolutional layers are generally too large to be accommodated on an FPGA platform. {\em Loop Tiling} is employed to partition the iteration space of the loop into smaller tiles so that the working data set of a tile can be easily accommodated using the available on-chip BRAM storage in FPGAs. Listing~\ref{lst:tiled_mm} shows the tiled matrix multiplication after {\em Loop Tiling}. The portion highlighted (Line 5-14) is the key computation of a tile and we accelerate this portion with FPGA-based PEs (explained in Section~\ref{subsubsec:hw_accelerator}) and NEON cores.

\begin{figure}[t!]
	\centering
	\includegraphics[width=0.7\columnwidth]{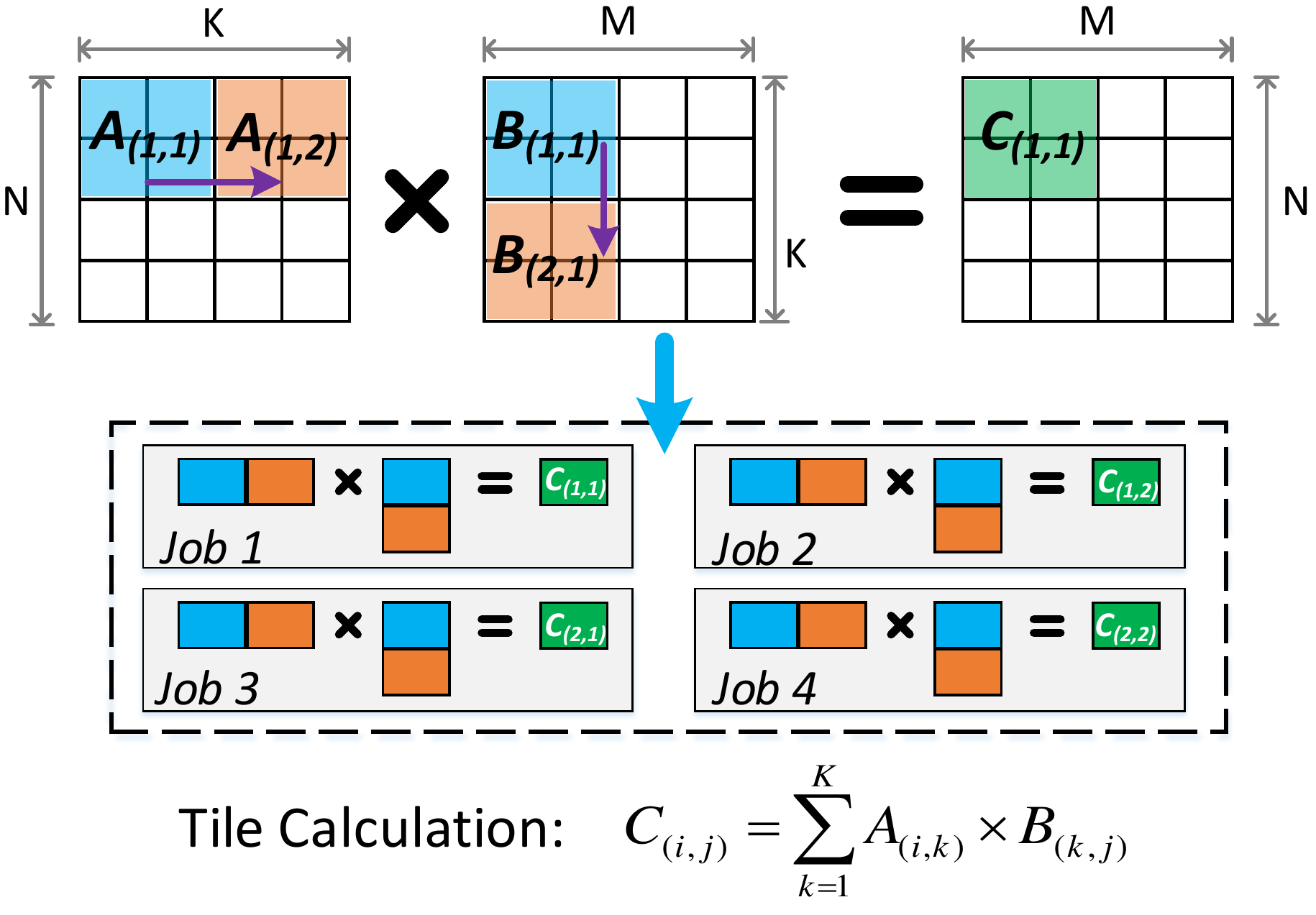}
	%\vspace{0.05em}
	\caption{Job: Workload Granularity of a Tiled MM}
	%\vspace{-1.5em}
	\label{fig:tiled_mm}
\end{figure}

{\bf Workload Granularity and Computation}: Figure~\ref{fig:tiled_mm} shows a tiled MM example with $2 \times 2$ tile size. In {\em Synergy}, the workload granularity of a tiled MM computation is called a {\it job}, which is defined as the computation required to output a tile, $C_{(i,j)}$, of an output feature map $C$. A {\it job} is a structure as shown in Listing~\ref{lst:job_argument} containing the base addresses of the arrays (A, B and C), the input data dimensions (m, n and k), the tile index and the layer ID, which is used to identify the CONV layer that owns the job. Each CONV layer generates a set of jobs. In a CONV thread, we implement a {\it courier} function that sends the jobs to the accelerators (PEs and NEONs). When an accelerator gets a job, it first calculates the memory addresses of the required tiles of input/output feature maps with the base address, data dimension and tile index provided by the job, and fetches the tiles from the external DDR memory to its local storage with the memory controller. After computation is completed, the PE stores the output tile back to the DDR.

\begin{lstlisting}[language=C, caption=The Structure of Job, basicstyle=\tiny, label={lst:job_argument}]
typedef struct {
/*The base addresses of input and output feature maps*/
DATA_TYPE A_addr; DATA_TYPE B_addr; DATA_TYPE C_addr;
/* Data dimension of input/output feature maps */
DATA_TYPE m;      DATA_TYPE n;      DATA_TYPE k;
/* Index used to locate the tile */
DATA_TYPE t1;     DATA_TYPE t2;
DATA_TYPE layer_id; /* To track the CONV layer */
} job_t;
\end{lstlisting}

{\bf Heterogeneous Accelerators}: As we target Xilinx Zynq SoC, {\em Synergy} uses the FPGA-based PEs and two NEON cores in the ARM A9 processor as the accelerators. A PE is an FPGA implementation of tiled MM. PEs can have different optimizations, and thus performance of PEs might be different. Number of PEs is dependent on the available resource in the target FPGA device. We explain the PE design in Section~\ref{subsubsec:hw_accelerator} in more detail. To leverage the NEON cores, we have implemented the MM kernel in NEON assembly code. This assembly code is encapsulated in two separate software threads, one corresponding to each NEON core, creating two NEON accelerators.

{\bf Accelerator Clusters}: From the software perspective, {\em Synergy} groups the heterogeneous accelerators into clusters. For example, in Figure~\ref{fig:sw_flow}, Cluster-0 has two NEON cores and two FPGA-based PEs, while Cluster-1 groups three PEs. Each cluster has a private workload pool, {\it Job Queue}, as shown in Figure~\ref{fig:sw_flow}. A {\it job queue} is a synchronous buffer, storing the address of the jobs. Each CONV layer is assigned to a cluster by default.  Different CONV layers can share the same cluster, for example CONV-0 and CONV-1 are mapped to Cluster-0 and CONV-2 uses Cluster-1. Mapping of CONV layers and clusters is decided by the number of jobs a CONV layer has. A CONV layer with less workload will be mapped onto a less powerful cluster and vice-versa. In addition, a designer can define the number of clusters and the corresponding accelerator combinations simply with a hardware configuration file as shown in Figure~\ref{fig:synergy_hw_generator}. In this case, the hardware accelerators will be synthesized and the required hardware-software interface will be automatically generated in the {\em Synergy} framework (see Section~\ref{subsubsec:generator}).

The CONV layers assigned to a cluster send their jobs to the {\it Job Queue} and use all the available accelerators in the cluster. Once the cluster detects jobs in the {\it job queue}, it dispatches the jobs to the synchronous buffers attached to each accelerator. Then, the accelerators work on the jobs and inform the cluster when they finish.

\subsubsection{\bf Delegate Threads}
\label{subsubsec:delegate_thread}
To abstract away the hardware accelerators, we deploy {\it delegate threads} introduced in ReconOS~\cite{ReconOS}. A delegate thread is a software wrapper for an FPGA-based accelerator, which can execute operating system (OS) calls on behalf of the associated accelerator. From the OS perspective, the delegate threads are software threads and can interact with the traditional software threads.

In {\em Synergy}, a delegate thread is created corresponding to each FPGA-based PE. Once launched, it initializes the hardware system and sends start signal to the associated accelerator via the first-in-first-out (FIFO) control buffer shown in Figure~\ref{synergy_hw_arch}. Then, the delegate thread waits for a request from the accelerator to execute a job. When an accelerator sends a {\it job} request, the delegate thread obtains the address of the job from its associated cluster and sends back to the accelerator, waiting for the accelerator's acknowledgment. Upon receiving the address of the job, the accelerator obtains the contents of a job structure, fetches the tile data of input arrays via the memory controller and performs the MM calculations. Once it finishes, the accelerator issues a signal to the delegate thread and acknowledges the completion of the tile calculation. The delegate thread repeats the above steps until all the jobs are finished.

\subsubsection{\bf Self-balancing: Work Stealing Scheduler}
%\todoinline{Need to mention why we need this}

\label{subsec:stealing}
\begin{figure}[t]
	\centering
	%\vspace{-0.5em}
	\includegraphics[width=0.7\columnwidth]{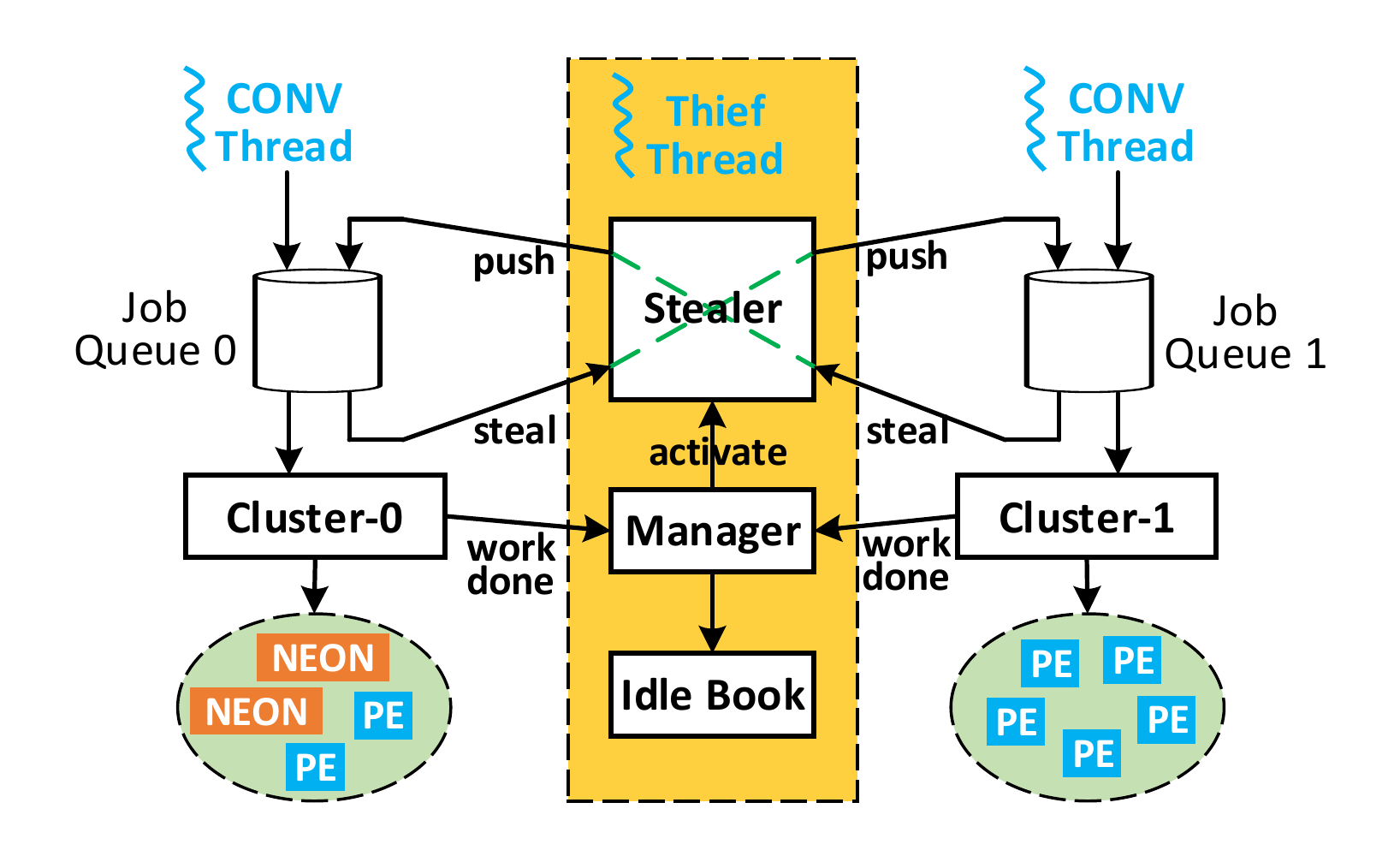}
	\caption{Work Stealing Execution Flow}
	%\vspace{-1.0em}
	\label{fig:work_stealing}
\end{figure}

{\em Synergy} clusters the FPGA-based PEs and the NEON accelerators into multiple clusters, so that the threads corresponding to multiple CONV layers can execute concurrently achieving better throughput. This approach also increases the accelerator utilization. However, as the workload of CONV layers varies depending on the data dimensions, an improper cluster configuration may lead to workload imbalance among the clusters. Some clusters might complete their workload early and stay idle, wasting precious computing resources. Therefore, clusters should be carefully partitioned and statically mapped to different CONV layers, so that the runtime of each cluster spent on processing the associated workload is balanced~\cite{Shen_RP_CNN_ISCA}. This can increase the accelerator utilization and improve the performance. However, finding the optimal cluster configuration is not easy. It requires profiling the performance of different cluster combinations for the input data dimensions of the CONV layers for the specific CNN model and perform a detailed design space exploration to identify the best cluster configuration for static mapping. Then the identified clusters and PEs have to be synthesized on the FPGA. However, this approach is challenging and time-consuming, especially without extensive FPGA expertise. In {\em Synergy}, we introduce dynamic workload balancing technique, work-stealing, to bypass this optimization problem.

%In Synergy, we try to ease the pressure of seeking the best cluster configuration by introducing dynamic workload balancing technique, work stealing, on this software/hardware multi-threading design.

%Therefore, to achieve the best accelerator utilization and improve performance, clusters should be carefully partitioned and statically mapped to different CONV layers.

%To find the best cluster configuration, one method is to use the brute-force search approach by profiling performance of different cluster combinations for various input data dimensions and suggest the best cluster configuration for static mapping. In Synergy, we try to ease the pressure of seeking the best cluster configuration by introducing dynamic workload balancing technique, work stealing, on this software/hardware multi-threading design.

This self-balancing technique is based on the job granularity and does not require the best cluster configuration as the idle cluster can steal jobs from the busy clusters. {\em Synergy} enables work stealing by introducing a thief thread. The thief thread consists of a {\it manager}, {\it idle book} and {\it stealer}. The {\it manager} checks the status (idle or busy) of the clusters and activates the {\it stealer} if necessary. The {\it idle book} records IDs of the idle clusters, while the {\em stealer} can steal jobs from the victim clusters and push these jobs to the idle clusters. Figure~\ref{fig:work_stealing} shows the work-stealing flow. Initially, {\em Synergy} dispatches the jobs of different CONV layers to job queues of different clusters. Due to the workload imbalance of the CONV layers, some clusters may finish the assigned workload earlier and remain idle. Let us assume that {\it Cluster-0} finishes first and {\it Cluster-1} is still busy. {\it Cluster-0} then notifies the {\it manager} of the thief thread, as its work has been done. The {\it manager} records {\it Cluster-0} in the {\it idle book} and activates the {\it stealer}. After activation, the {\it stealer} tries to steal jobs from the clusters that are not in the {\it idle book}. Once it succeeds, the {\it stealer} dispatches the jobs to the idle clusters and the {\it manager} removes the clusters from the {\it idle book}. In this manner, {\em Synergy} can fully utilize the accelerator resources and achieve load balancing. Different from the static mapping technique, the work-stealing approach does not rely on any specific cluster configuration to achieve workload balance. It eases the pressure of seeking the best cluster configuration and does not require designer's effort.
	
\subsubsection{Other Layers and Preprocessing functions}
\label{subsubsec:other_layers}
%\todoinline{Briefly explain pooling, activation, FC layers and some preprocessing functions, eg., im2col, normalization ...}
A CNN contains many other layers, which are executed by the ARM CPU cores in the {\em Synergy} framework. {\it Fully connected (FC) Layer}: This layer is usually used at the end of a network to compute the class scores, resulting in as many outputs as there are classes. {\it Pooling layer}: This layer progressively reduces the spatial size of the output from the previous layer to reduce the amount of parameters and computation in the network. {\it Activation layer}: This layer comprises of a non-linear function that does a 1-to-1 mapping of each of the outputs from the previous layer to an activation value. {\em Synergy} supports all kinds of activation functions.

A CNN also contains a few preprocessing functions within the layers such as {\it im2col} and {\it normalization} that take non-negligible time on embedded CPUs. {\it im2col} (mentioned in Section~\ref{subsubsec:conv_layer}) is used for data layout transformation. {\it Normalization} is used to scale all the input data to values between 0 and 1 during the inference phase. The overheads of these sequential portions are partially hidden by HW/SW multi-threaded pipeline in {\em Synergy}.

%% file: hardwareflow.tex
\subsection{Hardware Architecture}
\begin{figure}[t!]
	\centering
	%\vspace{-0.5em}
	\includegraphics[width=0.9\columnwidth]{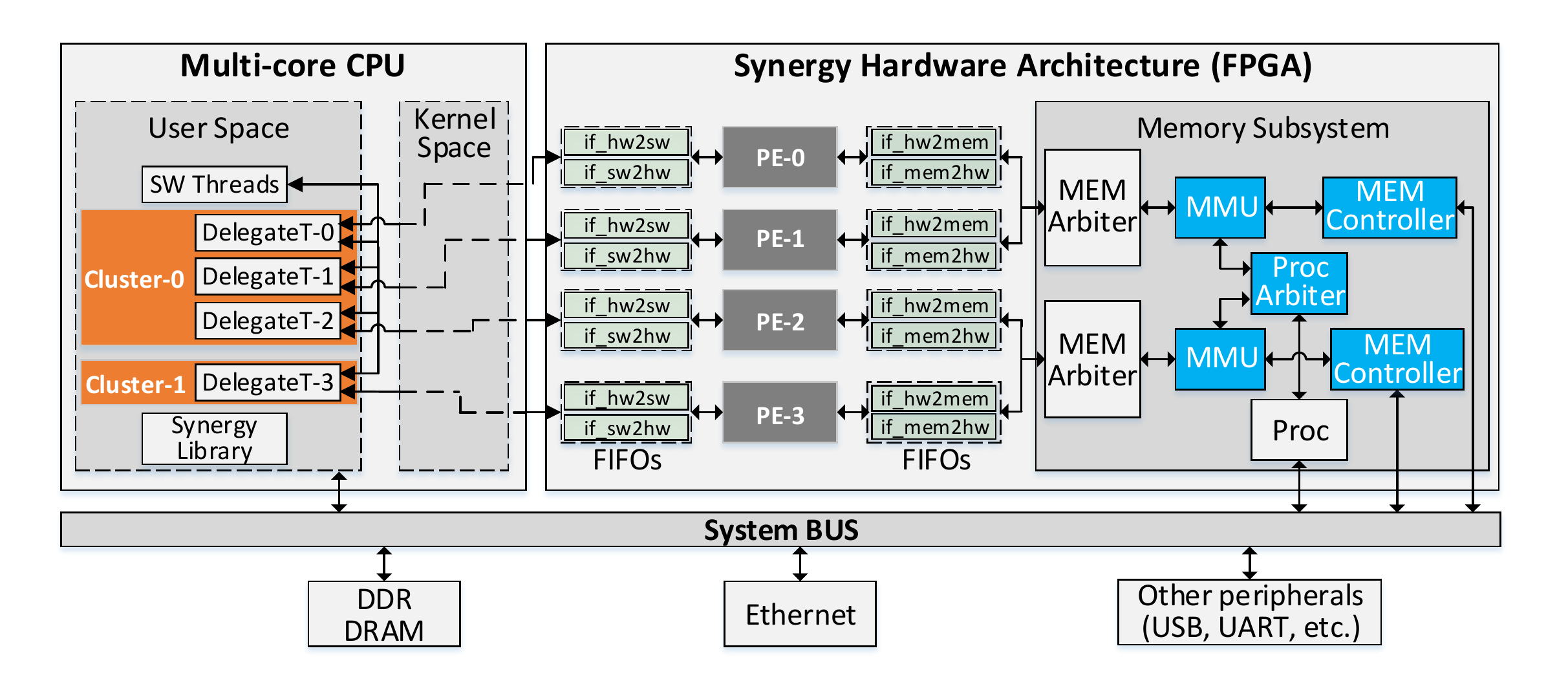}
	\caption{The Hardware Architecture}
	%\vspace{-1em}
	\label{synergy_hw_arch}
\end{figure}

Figure~\ref{synergy_hw_arch} shows an example {\em Synergy} hardware architecture example with four FPGA-based PEs. The architecture is adapted from ReconOS~\cite{ReconOS}. In this architecture, the software communicates with the hardware accelerators via control FIFOs ({\it if\_hw2sw} and {\it if\_sw2hw}). At the software side, a delegate thread ({\it DelegateT}) interacts with other software threads on behalf of the associated PE. Data transactions of a PE are handled by the Memory Subsystem via two FIFOs ({\it if\_hw2mem} and {\it if\_mem2hw}). In the following subsections, we discuss the accelerator design, memory subsystem, and the hardware architecture generator.

\subsubsection{Accelerator Design}
\label{subsubsec:hw_accelerator}
\begin{sloppypar}
%\todoinline{1. Mention challenges:a. large input data -> need loop tiling and properly select its tile size so that we do not degrade effeciency of data reuse; b. computation/communication optimizations on MM accelerators; c. latency of accelerator and memory transaction should be balanced;}
As mentioned earlier, {\em Synergy} processes CONV layers as matrix multiplication (MM) operations accelerated using NEON cores and FPGA-based PEs. In this section, we mainly focus on FPGA-based accelerators and discuss several design challenges. The FPGA-based accelerator for MM is the processing engine (PE) shown in Figure~\ref{synergy_hw_arch}, which is generated by a commercial high-level synthesis (HLS) tool, Vivado HLS~\cite{Xilinx}. HLS is used to convert algorithms in high-level specification (i.e., C/C++) into hardware languages (VHDL/Verilog). It provides optimization options, a.k.a pragmas, such as loop unrolling, array partitioning and pipelining, to explore diverse hardware architecture with different area and performance tradeoff.

As mentioned in Section~\ref{subsubsec:conv_layer}, due to the large input size of MM in CONV layers, we deploy {\em Loop Tiling} on MM and partition the iteration space of the loop into smaller tiles so that data size of a tile can be easily accommodated on available BRAM. {\em Loop Tiling} exposes potential parallelism of MM as different tiles are independent. We exploit the parallelism by instantiating multiple PEs under FPGA resource budget to process the tiles simultaneously, while exploring hardware architectures of a PE with HLS pragmas. Opening up more parallelism per PE limits the number of PEs that can be accommodated on FPGA due to resource constraints~\cite{mpseeker}.

%\begin{lstlisting}[language=C, caption=Code snippet of {\em tiled\_thread}, basicstyle=\tiny, label={lst:tiled_thread}, linebackgroundcolor={\ifnum\value{lstnumber}>14
%\ifnum\value{lstnumber}<17\color{yellow}\fi\fi}]
\begin{lstlisting}[language=C, caption=Pseudo Code for the HLS Template of a {\em PE}, basicstyle=\tiny, label={lst:hls_pe}, escapechar=|]
ProcessingEngine(if_sw2hw , if_hw2sw,
                 if_hw2mem, if_mem2hw) {
/* Simplified pragmas */
#pragma interface ap_fifo port=if_sw2hw
#pragma interface ap_fifo port=if_hw2sw
#pragma interface ap_fifo port=if_hw2mem
#pragma interface ap_fifo port=if_mem2hw
#pragma interface ap_ctrl_none port=return
  ...
  wait_for_start_signal(); |\label{line:start}|
  job_t job;
  while(1) {
    uint32 job_address = ask_for_a_job(); |\label{line:ask_job}|
    job = read_job(job_address); |\label{line:get_job}|
    parse_job(job, &Aaddr,&Baddr,&Caddr, |\label{line:parse_job1}|
              &m,&n,&k, &t1,&t2,&layerID); |\label{line:parse_job2}|
    tiled_mm(Aaddr,Baddr,Caddr,m,n,k,t1,t2, |\label{line:compute1}|
             if_hw2mem, if_mem2hw); |\label{line:compute2}|
    send_acknowledgment(layerID); |\label{line:acknowledgment}| }
}
\end{lstlisting}
%\vspace{-1em}
%we apply loop pipelining pragma at the second loop level of matrix multiplication in {\em mm\_kernel\_comp()}.

%As we target low-end embedded platforms, FPGA resources are quite limited and we can not generate a dedicated accelerator for each CONV layer in a CNN model like~\cite{going_deeper_cnn}\cite{Zhang_FPGA15_cnn}\cite{Caffeine_iccad2016}\cite{cnn_multi_fpga}. Therefore, in our implementation, we reuse the HW-IPs for each CONV layer in a CNN model. Due to various loop bounds of MM kernels in different CONV layers, we include border detection in data transaction (copy/send) functions.
\end{sloppypar}
\vspace{0.3em}
{\bf Processing Engine (PE)}: The pseudo code for the HLS template in Listing~\ref{lst:hls_pe} demonstrates the general execution flow of a PE. A PE interacts with its associated delegate thread in the user space via control FIFOs ({\em if\_hw2sw} and {\em if\_sw2hw}). For data transaction, the PE cooperates with the {\it Memory Subsystem} (Section~\ref{subsubsec:memory}) through memory FIFOs ({\em if\_hw2mem} and {\em if\_mem2hw}). At the beginning, the PE waits for a start signal issued from its associated delegate thread. Line~\ref{line:ask_job} - \ref{line:acknowledgment} in Listing~\ref{lst:hls_pe} shows the logic to compute a job. The PE first acquires a job by sending requests to the delegate thread. The real computation of MM is performed in {\it tiled\_mm}. The skeleton of {\it tiled\_mm} is shown in (Line~\ref{line:tiledMM1}-\ref{line:tiledMM2}) in Listing~\ref{lst:tiled_mm}. The {\it mm\_tile} function can be summarized as the following four steps: \protect\circled{1} It computes locations of tiles required of the input arrays ($A$ and $B$) in the main memory; \protect\circled{2} It then fetches a tile of data  to local memory ($a$ and $b$); \protect\circled{3} It performs matrix multiplication and adds the partial result with a local array $c$; \protect\circled{4} {\em mm\_tile} repeats Step 1 until it exhausts a row of $A$ and a column of $B$; \protect\circled{5} {\em mm\_tile} stores the output data back to the main memory. An acknowledgment will be sent to the delegate thread once the PE finishes computation.

{\bf Computation optimizations in {\em mm\_tile}}: Loop pipelining is a crucial optimization option provided by HLS. As the technique can overlap the execution of operations from different iterations, it has great impact on system throughput. Throughput of a loop depends on the initiation interval ($II$), which is defined as the number of cycles between consecutive initiations of the loop. In this work, we apply loop pipelining at {\em loop2} in Listing~\ref{lst:tiled_mm}. With the optimization, the HLS tool merges {\em loop1} and {\em loop2} into a new loop with larger loop bound ($newBound = TS*TS$) and completely unrolls the innermost loop ($loop3$). We define $lat_{loop3}$ as the latency of $loop3$. Then the latency $lat_{kernel}$ of the nested loop for kernel computation is calculated as $lat_{kernel} = (newBound-1)*II+lat_{loop3}$. When $newBound$ is large enough, $lat_{kernel}$ of the nested loop is decided by $II$.

As operations inside $loop3$ in Listing~\ref{lst:tiled_mm} have no data dependence, when $loop3$ is completely unrolled, operations in different iterations can be ideally executed in parallel. However, the parallelism is constrained by the memory bandwidth. Local buffers ($a$ and $b$) are implemented with FPGA BRAM resource. By default, a local buffer has only two read-ports and one write-port. Thus, when $loop3$ is completely unrolled, only two memory read requests to each buffer ($a$ and $b$) can be served, even if $TS$ read requests are generated. This makes $II$ to be $TS/2$ and limits performance of an accelerator. To improve $II$, we can leverage the array partitioning pragma to split the buffer into multiple banks where each bank has two read-ports and one write-port. With loop pipelining and array partitioning, the accelerator requires more multiplication and addition units, and thus more compute resources. Opening up more parallelism per PE limits the number of PEs that can be accommodated on FPGA due to resource constraints. Given a FPGA device, the tile size, the settings for HLS pragmas, and the number of PEs can be done automatically decided via design space exploration (DSE)~\cite{mpseeker}.

%In current Synergy implementation, we manually decide the number of accelerators and pragma combinations used in HLS and leave the DSE engine integration as our future plan.
%This might decrease number of accelerators can be instantiated due to available resources on the FPGA platform.

%Loop unrolling, loop pipelining and array partitioning are three prominent optimization pragmas in HLS that significantly affect performance and area consumption of an accelerator.

{\bf Communication optimization in {\em mm\_tile}}: For tiled matrix multiplication, {\em Synergy} overlaps the data transfer cost with the computation cost by leveraging double buffering, i.e., instantiating two buffers for each local array. This significantly improves the throughput of tiled MM.

{\bf Zero Padding in {\em mm\_tile}}: In {\em Synergy}, the hardware accelerators are shared among the convolutional layers. This implies the same MM accelerator of fixed size are used in different layers. As the loop bounds (or data dimensions) of MM in different convolutional layers are different, we may encounter scenarios where the fixed-size MM accelerator attempts to access out of the loop bound data of the input matrices or write data outside the bounds of the output matrix. Hence, we include border detection in {\em mm\_tile}. When fetching data, if the memory address exceeds the matrix border, the specific portion of the local buffer will be set to zero. Similarly, for writing data, {\em mm\_tile} ignores write requests if a memory address exceeds the given matrix borders.

\input{interface.tex}

%% file: interface.tex
\subsubsection{Memory Subsystem}
\label{subsubsec:memory}
%\todoinline{Briefly introduce how we transfer data}
The Memory Subsystem shown in Figure~\ref{synergy_hw_arch} is used to process memory requests from multiple PEs. It consists of memory arbiters ({\it MEM Arbiter}), memory management units ({\it MMUs}), memory controllers ({\it MEM Controllers}), a {\it Proc Arbiter} and a {\it Proc unit}. {\it MMU} is used to translate virtual addresses to physical addresses, while {\it MEM Arbiter} is employed to allocate memory transaction requests to the shared {\it MMU}. {\it Proc unit} is used to obtain the first-level translation page table address and handle page fault request, and {\it Proc Arbiter} allows multiple {\it MMUs} to access the {\it Proc unit}. {\it MEM Controllers} are implemented to access the DDR memory with AXI4 burst mode protocol. All the components in the Memory Subsystem are written in RTL code and constitute the {\it Hardware Template Library} as shown in Figure~\ref{fig:synergy_hw_generator}.

\begin{figure}[t]
	\centering
	\includegraphics[width=0.5\columnwidth]{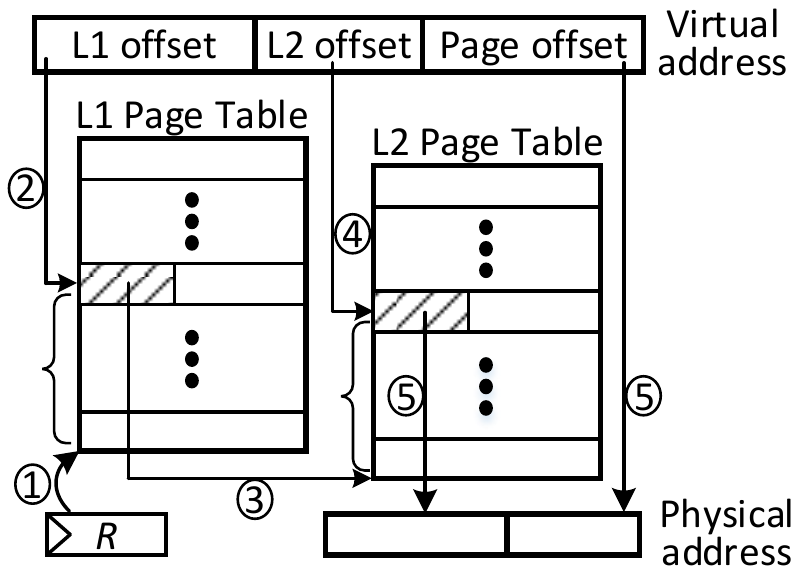}
	\caption{Virtual To Physical Address Translation~\cite{ARM}}
	%\vspace{-1em}
	\label{fig:v2p}
\end{figure}

{\bf Virtual to Physical Address Translation}: In a traditional HW/SW co-design approach, a device driver normally has a continuous memory address space in the Linux kernel. When a delegate thread tries to communicate with an FPGA PE, it first copies data from the user space to the allocated continuous memory (kernel space) in the device driver and sends the physical address of the memory to the PE. Then the PE obtains the data from the DDR memory via the {\it MEM Controllers}. In {\em Synergy}, we avoid the extra data copy in the acceleration of CONV layers. As mentioned in Section~\ref{subsubsec:delegate_thread} and~\ref{subsubsec:hw_accelerator}, a PE obtains an address of a job directly from the delegate thread in the user space and the job content includes the base memory address of input/output arrays in the user space. Those are virtual addresses. In ARM Cortex-A9 architecture~\cite{ARM}, virtual addresses are translated to physical addresses by a two-level (L1 and L2) page table walk as shown in Figure~\ref{fig:v2p}. The base address of the L1 page table is stored in a CPU system register {\it R}~\cite{ARM}, which can be accessed in the kernel space. {\em Synergy} supports this two-level page table walk in FPGA. During the FPGA initialization in {\em Synergy}, the {\it Proc unit} obtains the base address of the L1 page table via its device driver. Then, the {\it Memory Subsystem} translates the virtual address to physical address following the steps in Figure~\ref{fig:v2p}. In case of a page fault, the {\it Proc unit} triggers a CPU interrupt, obtains a new base address and repeats the translation process.

\begin{figure}[ht]
	\centering
	%\vspace{-0.5em}
	\begin{subfigure}[b]{0.47\linewidth}
		{
			\centering
			\includegraphics[width=0.9\columnwidth]{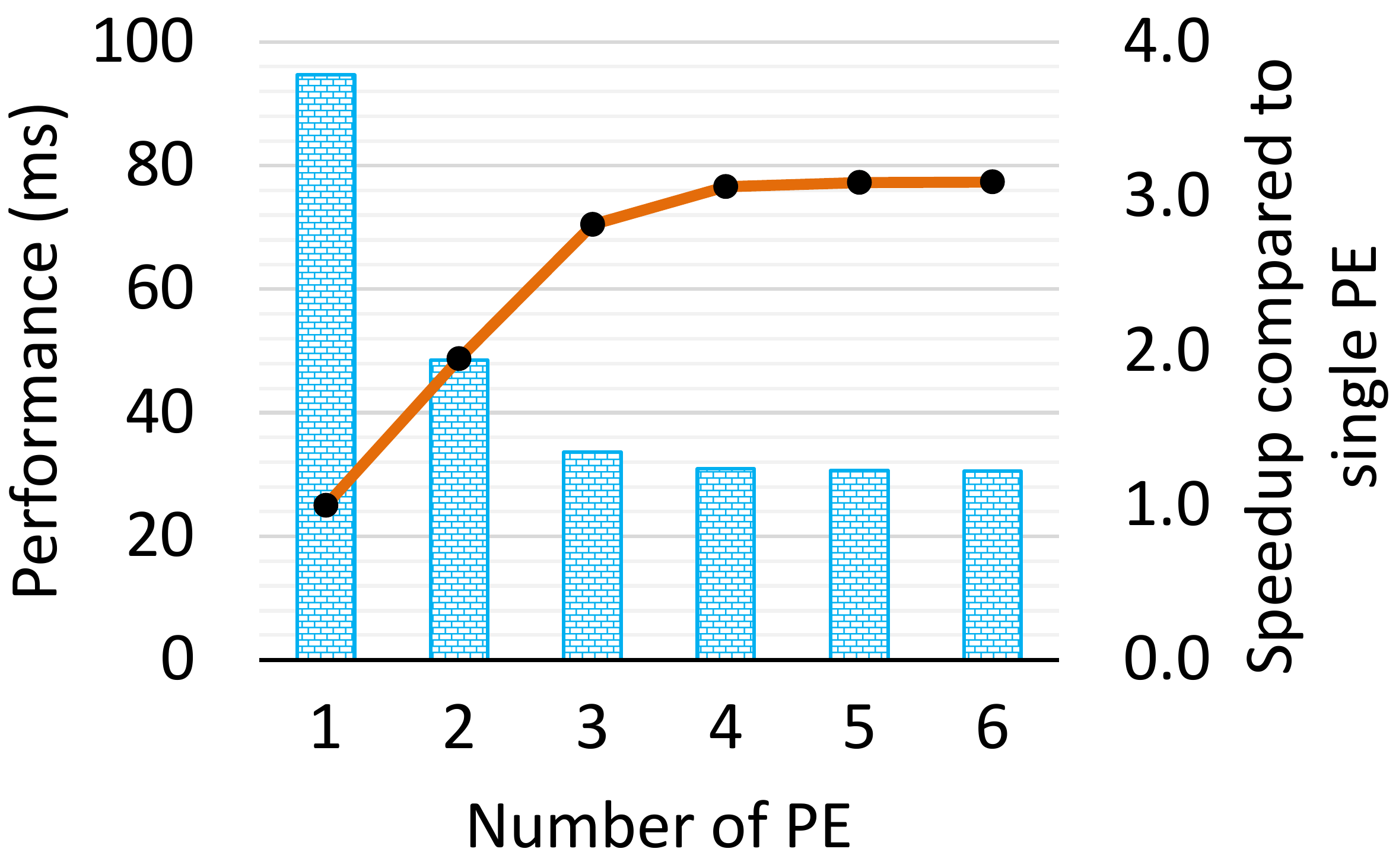}
			%\vspace{-1.5em}
			\caption{ReconOS}
			\label{fig:limit_sharedMEM}
		}
	\end{subfigure}
	\quad
	\begin{subfigure}[b]{0.47\linewidth}
		{
			\centering
			\includegraphics[width=0.9\columnwidth]{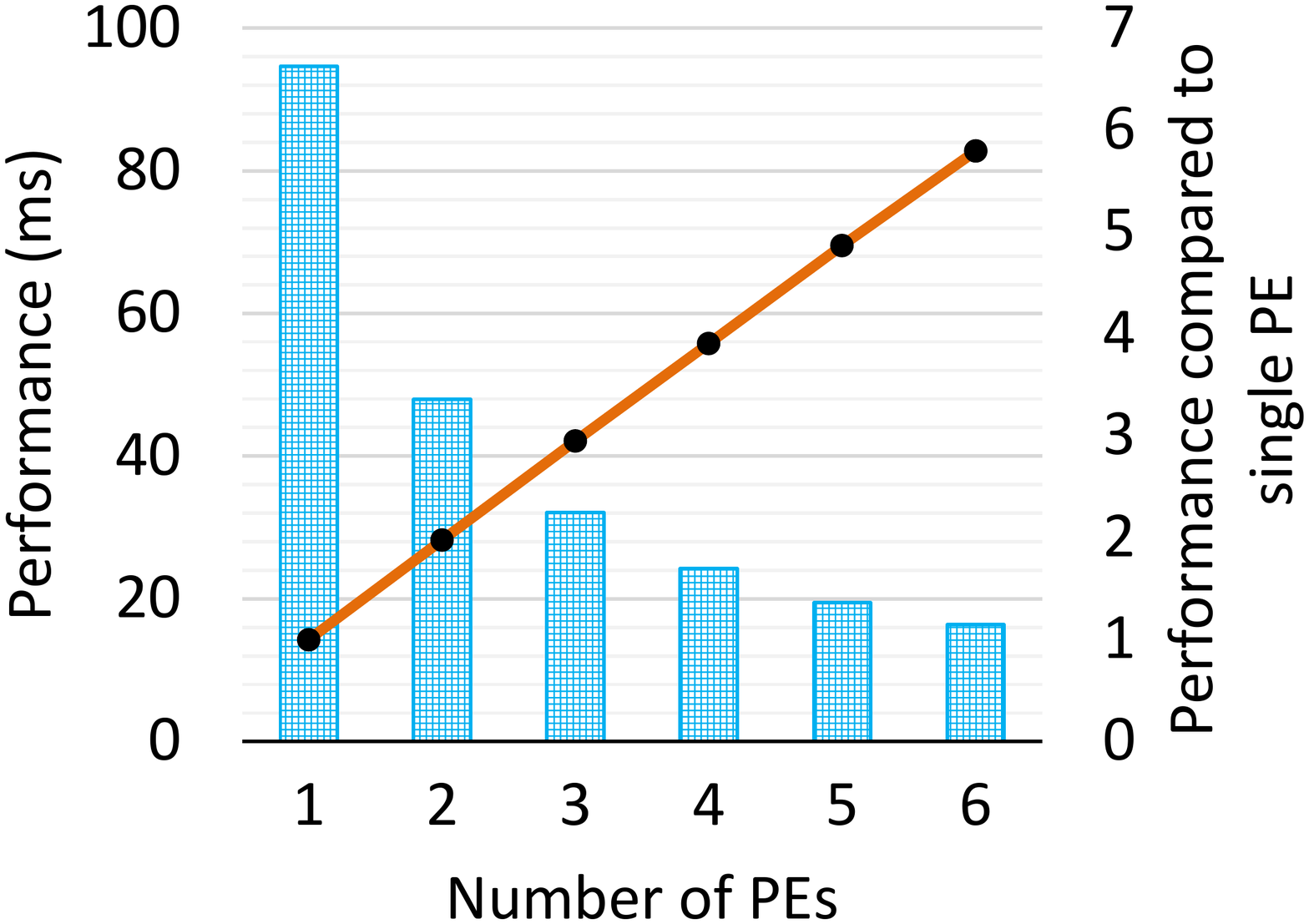}
			%\vspace{-1.5em}
			\caption{Synergy}
			\label{fig:multi_mem}
		}
	\end{subfigure}
	%\vspace{0.5em}
	\caption{Single-MMU vs. Multi-MMU Peformance}
	%\vspace{-1em}
	\label{fig:mmu_comp}
\end{figure}

%As based addresses are virtual addresses come from the delegate thread in user space, a memory management unit (MMU), which is a page-table-walk hardware~\cite{reconos_mmu}, in the Memory System is used to translate virtual addresses to physical addresses. Page faults from hardware accelerators are handled by the Proc unit. With physical addresses, the MEM Controller can access DDR DRAM (main memory) in single-word or burst mode.
%\todoinline{Explain why single MMU is not sufficient}
{\bf Multiple MMU Support}: ReconOS architecture~\cite{ReconOS} contains a single {\it MMU} and {\it MEM Controller}. The memory transactions from the PEs compete for the resources in the Memory Subsystem. As the number of PEs increases, the memory contention significantly degrades the system performance as shown in Figure~\ref{fig:limit_sharedMEM}. To solve the problem, {\em Synergy} instantiates multiple MMUs with at most two PEs sharing an MMU and MEM Controller. As the frequency of page faults is generally low in our case, multiple {\it MMUs} in {\em Synergy} share the same {\it Proc unit} via the arbiter logic {\it Proc\_Arbiter}.  Figure~\ref{fig:multi_mem} shows that the performance speedup increases linearly as we instantiate more PEs in {\em Synergy}.

%% file: selfbalancing.tex
\begin{figure}[t]
	\centering
	\includegraphics[width=0.6\columnwidth]{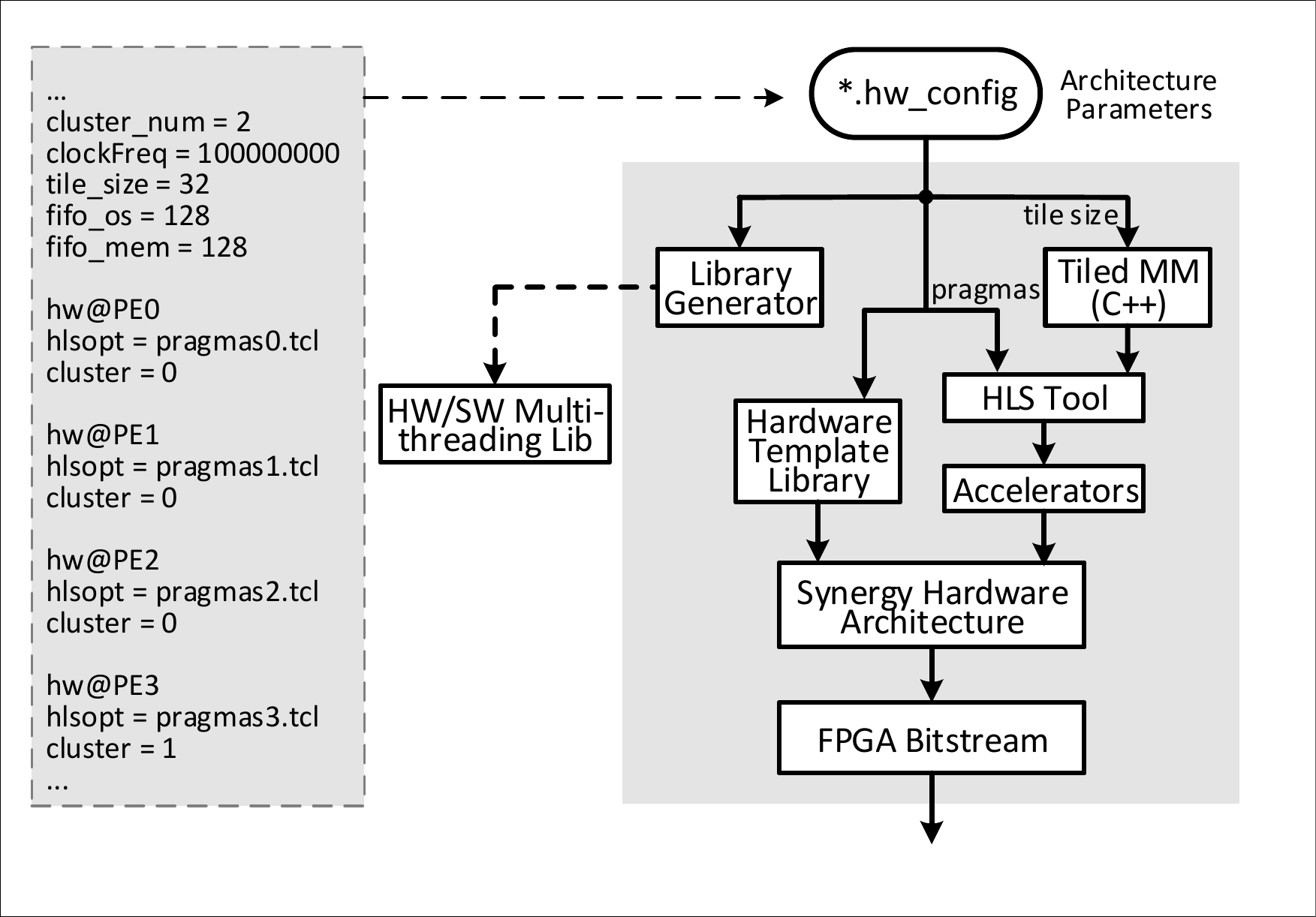}
	\caption{Hardware Architecture Generator}
	%\vspace{-1.5em}
	\label{fig:synergy_hw_generator}
\end{figure}

\subsection{Hardware Architecture Generator}
\label{subsubsec:generator}
%1. generally explain the Figure. 2. show format of hw.config and explain it;
{\em Synergy} provides a default accelerator architecture on a given FPGA device. However, for a new FPGA device or in case the developer is interested in customizing the accelerator architecture corresponding to a CNN model,
{\em Synergy} provides an architecture generator as shown in Figure~\ref{fig:synergy_hw_generator}. This automates the processes of generating PEs with HLS, the Hardware Architecture, and final FPGA bitstream. Input to the generator is a configuration file, {\em *.hw\_config}, containing the architecture parameters. The simplified format of this configuration file is shown in the left side of Figure~\ref{fig:synergy_hw_generator}, which creates the Hardware Architecture shown in Figure~\ref{synergy_hw_arch}. Moreover, based on the configuration file, the generator also compiles HW/SW multi-threading library ({\em Synergy} Library) to provide APIs required by Section~\ref{subsec:sw_comp}.

%% file: experiment.tex
\section{Experimental Evaluation}
\label{sec:results}
%\todoinline{1. Experimental Setup;\\2. HW-IP configuration (HLS pragma used, number of HW-IP, etc.);\\3. Benchmarks considered, also add a column to specify hw threads and HW-IPs combination}
%The Synergy framework is set up on a PC with an Intel Xeon CPU E5-2620 core at 2.10Hz with 64GB RAM, running Ubuntu 14.04 OS. We use Darknet~\cite{darknet13} as our deep learning package and extend it into multi-threaded version. The Synergy Hardware Architecture generation is modified based on ReconOS toolchain~\cite{ReconOS}\cite{ReconOS_hls}. As the current-version ReconOS can not support Vivado~\cite{Xilinx}, we use the Xilinx Embedded Development Kit~\cite{Xilinx} version 14.7 for FPGA bitstream generation. We are still working on porting Synergy to Vivado. For hardware accelerator generation, we leverage Vivado HLS version 2016.2. The Synergy hardware on FPGA is running at 100MHz. The low-end heterogeneous SoC platform used is Xilinx ZC702~\cite{Xilinx}. In this work, we consider 32-bit floating-point CNN models on the multi-threaded Darknet, and its accuracy is comparable to the standard CNN frameworks such as Caffe. All performance values are collected by an FPGA-based timer.

In this section, we evaluate the {\em Synergy} framework with multiple representative CNN models.

{\em Synergy} has been implemented on heterogeneous SoC platforms Zedboard~\cite{Xilinx} and Xilinx ZC702, both featuring the same Xilinx Zynq XC7Z020 device. Xilinx Zynq XC7Z020 is a low-end SoC in terms of its compute capability and the availability of limited FPGA resources. We report the performance and power numbers from the Xilinx ZC702 evaluation board because it has the power measurement support.  All performance results are collected using an FPGA-based timer.

We use {\em Darknet}~\cite{darknet13} as our deep learning package. {\em Darknet} is an open source neural network framework written in C. We use {\em Darknet} because it has a highly-optimized single-threaded software implementation and does not depend on any external library. We first compile {\em Darknet} for the ARM core in the Xilinx Zynq device.  Apart from the single-threaded software implementation, we create a multi-threaded pipelined version of {\em Darknet} to take advantage of inter-frame parallelism for high-throughput CNN implementation. The CPU-only implementations for various CNNs in this section are well-optimized and compiled by {\it gcc} with {\it -O3} optimization. As {\em Darknet} uses 32-bit floating-point CNN models, we also use 32-bit floating-point implementation both in software and hardware accelerators. The performance-power numbers of {\em Synergy} will improve substantially if 32-bit floating-point implementation is replaced by $N$-bit fixed-point implementation where $N << 32$. However, this optimization is orthogonal and complementary to our current approach. Even with floating-point, we achieve better throughput and energy-efficiency compared to contemporary fixed-point implementations on the same device.

We write assembly-language code to generate highly optimized NEON accelerators for the tiled matrix-multiplication operations. For hardware accelerator generation, we use Vivado Design Suite version 2016.2 for High-Level Synthesis (HLS). The tiled matrix multiplication code is written in C and are synthesized on FPGA using Vivado HLS with appropriate pragma settings as presented in Section~\ref{subsubsec:hw_accelerator}. {\em Synergy} uses two clusters (Cluster-0: 2 NEONs + 2 S-PE; Cluster-1: 6 F-PE) configuration across all benchmarks. The cluster configuration is chosen based on power/performance results across multiple CNNs and the {\em work stealing} technique can ensure that other CNN applications could work well on this fixed hardware architecture as well by balancing workload at runtime. The FPGA-based PEs run at 100MHz. The HW/SW multi-threading is implemented by adapting {\em ReconOS} open-source operating system for reconfigurable computing~\cite{ReconOS}\cite{ReconOS_hls}. The ARM cores run Linux, which is augmented with {\em ReconOS} to interface with the hardware accelerators. 

%We write assembly-language code to generate highly optimized NEON accelerators for the tiled matrix-multiplication operations. For hardware accelerator generation, we use Vivado Design Suite version 2016.2 for High-Level Synthesis (HLS). The tiled matrix multiplication code is written in C and are synthesized on FPGA using Vivado HLS with appropriate pragma settings as presented in Section~\ref{subsubsec:hw_accelerator}. {\em Synergy} uses two clusters (Cluster-0: 2 NEONs + 2 S-PE; Cluster-1: 6 F-PE) configuration across all benchmarks. The FPGA-based PEs run at 100MHz. The HW/SW multi-threading is implemented by adapting {\em ReconOS} open-source operating system for reconfigurable computing~\cite{ReconOS}\cite{ReconOS_hls}. The ARM cores run Linux, which is augmented with {\em ReconOS} to interface with the hardware accelerators. 

The entire {\em Synergy} framework is set up on a PC with an Intel Xeon CPU E5-2620 core running at 2.10Hz with 64GB RAM, running Ubuntu 14.04 OS. Given a CNN model, the {\em Synergy} framework is responsible to generate the appropriate software threads corresponding to the different layers of the network, interfacing the software threads with the delegate threads of the hardware accelerators, and creating a default mapping between the CONV layers and the accelerator clusters. The {\em Synergy} framework can also automate the FPGA bitstream generation given a hardware accelerator architecture configuration by the designer to customize {\em Synergy} implementation for a particular CNN model (if desired) or generate one for a new device.

%\todoinline{Add this one: There exist many popular open-source deep neural network frameworks that support CNNs such as Caffe~\cite{caffe} and Darknet~\cite{darknet13}. As Darknet does not depend on any external library and it is fairly easy to install it on embedded platforms, we extend it with multi-threaded support. }

%\todoinline{Add references in Table~\ref{tab:bench_arch}}
%\todoinline{Change the table, no need to show the network details, can just show how many conv layers, etc.}
{\bf Benchmarks}: Table~\ref{tab:bench_arch} shows seven CNN models used in this work and trained with {\em Darknet}.

\begin{table}[t]
	\centering
	\caption{Network Architectures of Benchmark CNN Models}
	\label{tab:bench_arch}
	%\resizebox{0.7\columnwidth}{!}{
		\begin{tabular}{|l|c|c|c|}
			\hline
			Benchmark & \begin{tabular}[c]{@{}c@{}}CONV \\ Layers\end{tabular} & \begin{tabular}[c]{@{}c@{}}Num. of \\ Layers\end{tabular}  & Description \\ \hline
			CIFAR\_Darknet~\cite{darknet13} & 4 & 9 & Object Recognition \\ \hline
			CIFAR\_Alex~\cite{cifar_alex} & 3 & 8 & Object Recognition \\ \hline
			CIFAR\_Alex+~\cite{cifar_alex} & 3 & 9 & Object Recognition \\ \hline
			CIFAR\_full~\cite{caffe} & 3 & 9 & Object Recognition \\ \hline
			MNIST~\cite{LeNet5} & 2 & 7 & Digit Recognition \\ \hline
			SVHN~\cite{svhn_bench} & 3 & 8 & Digit Recognition \\ \hline
			MPCNN~\cite{mpcnn_bench} & 3 & 9 & Gesture Recognition \\ \hline
		\end{tabular}
	%}
	%\vspace{-1em}
\end{table}

\subsection{Synergy Throughput and Energy-Efficiency}
\label{subsec:overall_perf}

\begin{figure}[h]
	\centering
	%\vspace{-0.8em}
	%\includegraphics[width=0.8\columnwidth]{./graphs/perf_mc_vs_ws.pdf}
	\includegraphics[width=0.6\columnwidth]{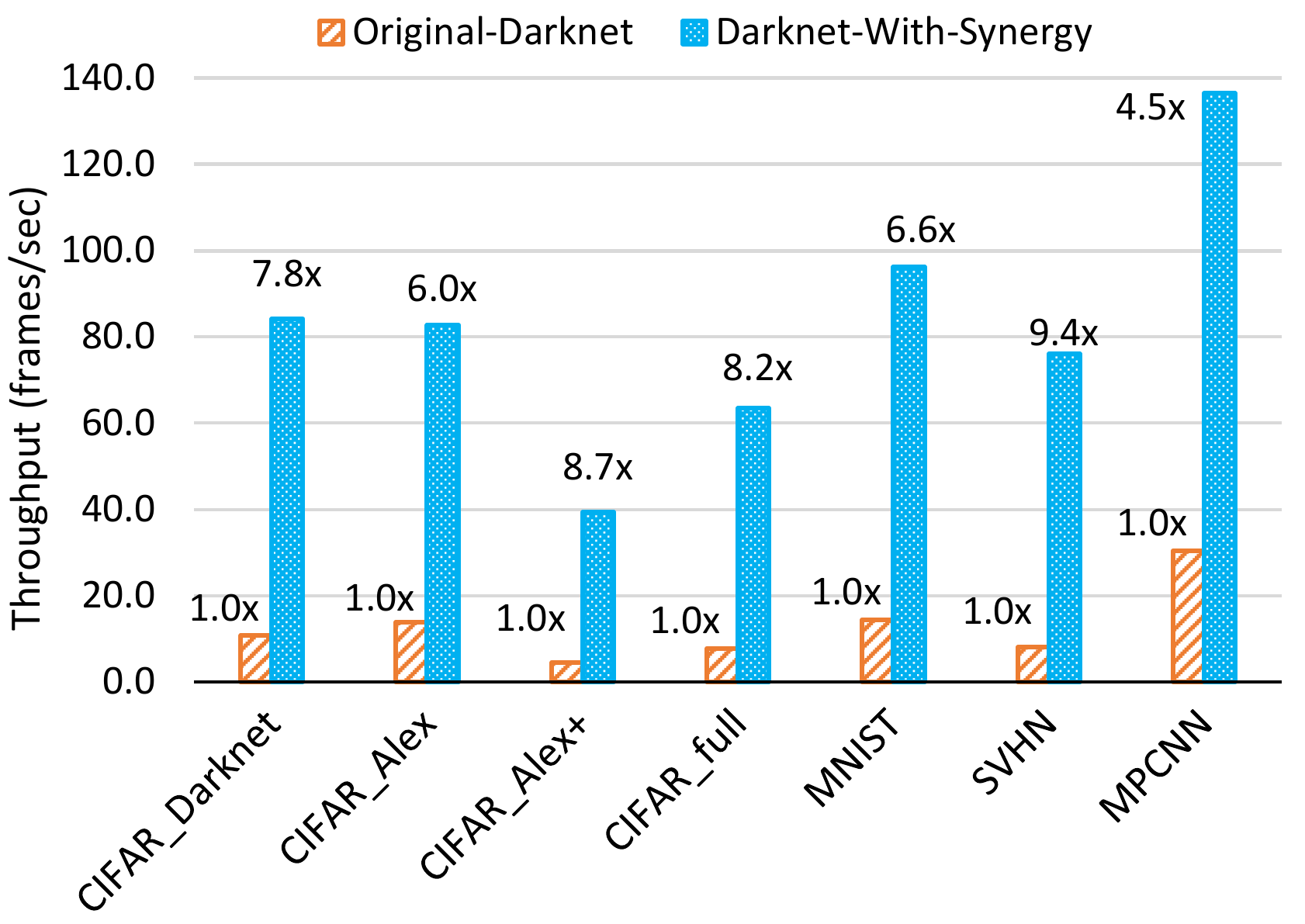}
	%\vspace{0.05em}
	\caption{Throughput improvement using Synergy}
	%\vspace{-1em}
	\label{fig:overall_throughput}
\end{figure}

{\bf Throughput}: Compared with the original single-threaded {\em Darknet} implementation running on ARM core, {\it Synergy} achieves average 7.3x throughput improvement as shown in Figure~\ref{fig:overall_throughput}.

\begin{figure}[h]
	\centering
	%\vspace{-0.5em}
	\includegraphics[width=0.6\columnwidth]{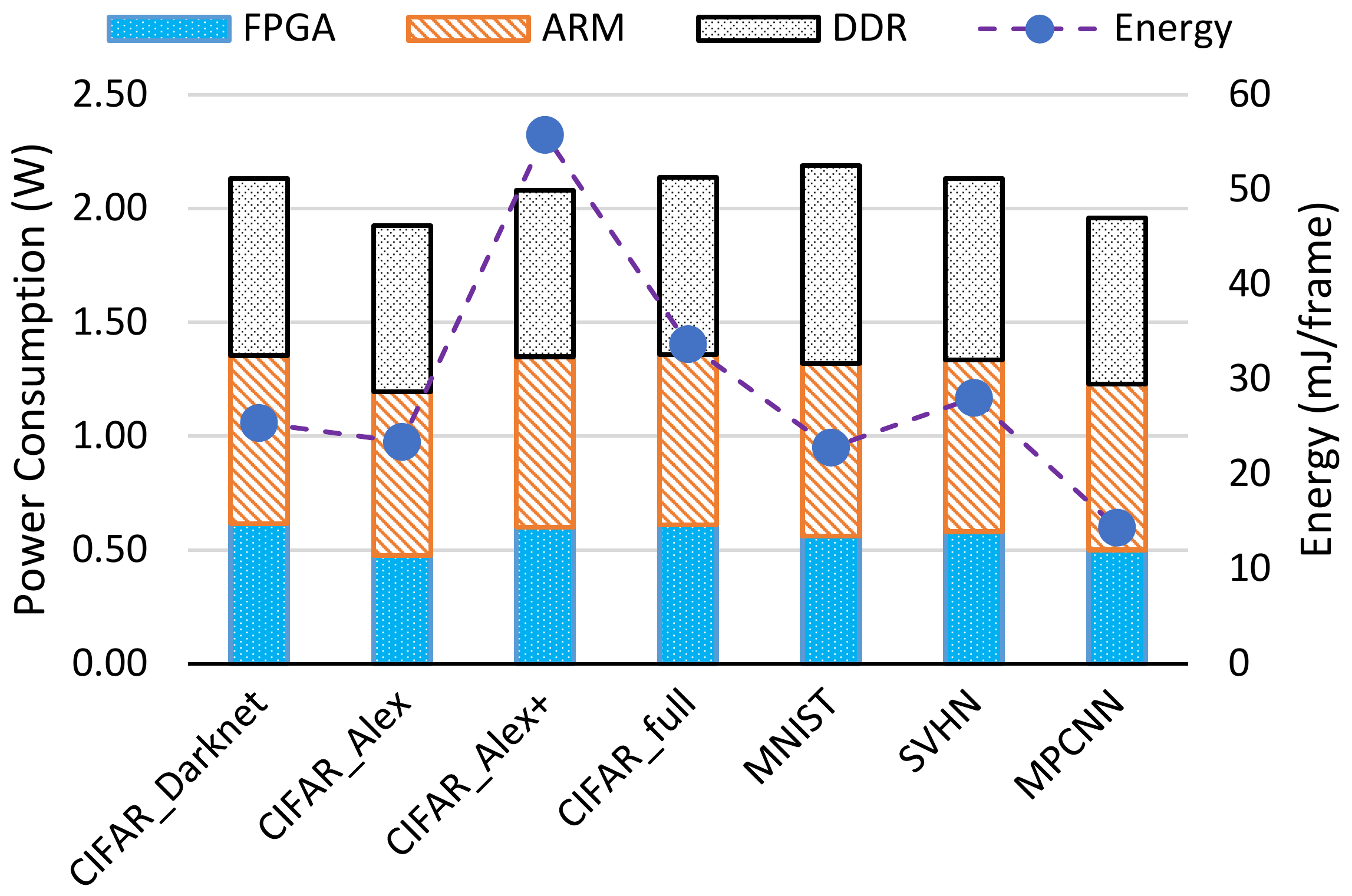}
	\caption{Power Distribution and Energy Consumption}
	%\vspace{-1em}
	\label{fig:power_energy}
\end{figure}

{\bf Power and Energy Consumption}: Figure~\ref{fig:power_energy} depicts the power distribution and energy consumption of {\em Synergy} system. The FPGA logic accounts for only 27\% of the total power consumption (around 2.08 W) averaged across all CNN models. The ARM cores and the DDR memory account for most of the power consumption. Compared with the power (1.52 W on average) measured for the CPU+NEON only implementations, {\it Synergy} incurs 36.63\% more power consumption.

\begin{table}[ht]
	\centering
	\caption{Energy and Performance per Watt Comparison: Original Darknet Versus Synergy}
	\label{tab:power-energy}
	\begin{tabular}{|l|c|c|c|c|c|c|}
		\hline
		\multirow{2}{*}{Benchmarks} & \multicolumn{3}{c|}{Energy (mJ/frame)} & \multicolumn{3}{c|}{Performance per watt (GOPS/W)} \\ \cline{2-7} 
		& Original       & Synergy   & Reduction (\%)    & Original  & Synergy  & Speedup  \\ \hline
		CIFAR\_Darknet              & 142.18           & 25.36   &  -82.16     & 0.14    &  0.80   &   5.61x         \\ \hline
		CIFAR\_Alex                 & 105.03           & 23.43   &  -77.70     & 0.16    &  0.80   &  4.48x          \\ \hline
		CIFAR\_Alex+                & 326.62           & 55.81   &  -82.91     & 0.16    &  0.70   &   5.85x         \\ \hline
		CIFAR\_full                 & 196.41           & 33.71   &  -82.84     & 0.13    &  0.94   &   5.83x         \\ \hline
		MNIST                       & 112.90           & 22.78   &  -79.83     & 0.20    &  0.78   &   4.96x         \\ \hline
		SVHN                        & 193.67           & 28.07   &  -85.50     & 0.14    &  0.98   &   6.90x         \\ \hline
		MPCNN                       & 47.87           & 14.37   &  -69.99     &  0.20    &  0.68   &   3.33x         \\ \hline \hline
		\multicolumn{3}{|l|}{mean}                       &    -80.13           & \multicolumn{2}{c|}{} &   5.28x         \\ \hline
	\end{tabular}
\end{table}

%\begin{table}[ht]
%	\centering
%	\caption{Power and Energy Comparison: Original Darknet Implementation Versus Synergy Design}
%	\label{tab:power-energy}
%	\begin{tabular}{|l|c|c|c|c|c|c|}
%		\hline
%		\multirow{2}{*}{Benchmarks} & \multicolumn{3}{c|}{Power (W)} & \multicolumn{3}{c|}{Energy (mJ/frame)} \\ \cline{2-7} 
%		& Original       & Synergy   & Increase (\%)    & Original  & Synergy  & Reduction (\%)  \\ \hline
%		CIFAR\_Darknet              & 1.53           & 2.14   &  39.54     & 142.18    & 25.36    & 82.16           \\ \hline
%		CIFAR\_Alex                 & 1.46           & 1.93   &  31.85     & 105.03    & 23.43    & 77.70           \\ \hline
%		CIFAR\_Alex+                & 1.48           & 2.08   &  40.54     & 326.62    & 55.81    & 82.91           \\ \hline
%		CIFAR\_full                 & 1.53           & 2.14   &  39.87     & 196.41    & 33.71    & 82.84           \\ \hline
%		MNIST                       & 1.64           & 2.19   &  33.54     & 112.90    & 22.78    & 79.83           \\ \hline
%		SVHN                        & 1.56           & 2.14   &  36.86     & 193.67    & 28.07    & 85.50           \\ \hline
%		MPCNN                       & 1.46           & 1.96   &  34.25     & 47.87     & 14.37    & 69.99           \\ \hline \hline
%		\multicolumn{3}{|l|}{mean}                       &    36.63           & \multicolumn{2}{c|}{} & 80.13           \\ \hline
%	\end{tabular}
%\end{table}

Table~\ref{tab:power-energy} shows the energy and performance per watt comparison between the original single-threaded {\em Darknet} implementation running on ARM cores and the {\em Synergy} design. Considering power consumption, the {\em Synergy} design consumes 36.63\% more power on average, as it fully leverages the heterogeneity of the ZYNQ platform. Although the power consumption increases, the {\em Synergy} implementation achieves much higher throughput (7.3x speedup), and thus reduces 80.13\% energy consumption averaged across all CNN models compared to the original {\em Darknet} on ARM cores.

%\begin{table}[h]
%	\centering
%	\caption{Resource Utilization of Synergy on ZC702}
%	\label{tab:fpga_area}
%	%\resizebox{0.7\columnwidth}{!}{
%	\begin{tabular}{|l|c|c|c|c|}
%		\hline
%		& DSP  & BRAM18K & LUT   & FF     \\ \hline
%		Available & 220  & 280     & 53200 & 106400 \\ \hline
%		Ratio(\%) & 70.0 & 28.6    & 93.1  & 46.0   \\ \hline
%	\end{tabular}
%	%}
%	%\vspace{-1em}
%\end{table}

{\bf FPGA Resource Consumption}: Hardware accelerators generated by Vivado HLS have great impact on FPGA resource consumption. With the limited FPGA resource budget, opening up more parallelism via HLS pragmas reduces the number of hardware accelerators that can fit in Xilinx ZC702. Therefore, we explore diverse architectures of hardware accelerators by traversing different tile size and HLS pragma combinations consisting loop unrolling, loop pipelining and array partitioning. In this work, the tile size is set to be 32 based on empirical evaluation. On ZC702 device we instantiate 6 faster FPGA-based processing engines (F-PE) with loop pipelining pragma applied at $loop2$ in Listing~\ref{lst:tiled_mm} and 2 slower PE (S-PE) with loop unrolling (factor = 2) and loop pipelining at $loop3$. %Table~\ref{tab:fpga_area} shows the FPGA area consumption by the synthesized hardware architecture in {\em Synergy}. LUT is the resource bottleneck and limits the number of hardware accelerators that can be accommodated on ZC702.

\begin{table}[ht]
	\centering
	\caption{Comparison With Recent FPGA-based CNN Works. `*' indicates values estimated from charts}
	\label{tab:energy_comp}
	\resizebox{\columnwidth}{!}{
		\begin{tabular}{|l|c|c|c|c|c|c|c|c|c|}
			\hline
			& \multicolumn{2}{c|}{CaffePresso~\cite{caffepresso_cnn}}        & \multicolumn{2}{c|}{fpgaConvNet\cite{fpgaConvNet_fccm16}\cite{fpgaConvNet}}           & \multicolumn{2}{c|}{DeepBurning~\cite{DeepBurning}} & \multicolumn{3}{c|}{Synergy}               \\ \hline
			Device                                                           & \multicolumn{2}{c|}{7Z045}              & \multicolumn{2}{c|}{7Z020}              & \multicolumn{2}{c|}{7Z020}       & \multicolumn{3}{c|}{7Z020}                 \\ \hline
			Clock (MHz)                                                      & \multicolumn{2}{c|}{180}                & \multicolumn{2}{c|}{100}                & \multicolumn{2}{c|}{100}         & \multicolumn{3}{c|}{100}                   \\ \hline
			Precision                                                        & \multicolumn{2}{c|}{16-bit Fixed-point} & \multicolumn{2}{c|}{16-bit Fixed-point} & \multicolumn{2}{c|}{Fixed-point} & \multicolumn{3}{c|}{32-bit Floating-point} \\ \hline
			Benchmarks                                                       & MNIST              & CIFAR\_full        & MNIST     &   MPCNN       & MNIST        & CIFAR\_full       & MNIST             & CIFAR\_full     &   MPCNN       \\ \hline
			Latency(ms)                                                         & 16.0               & 28.0               & {\bf --}       &   {\bf --}       & 14.3         & 21.4              & 24.3              & 33.2       & 12.2            \\ \hline			
			\begin{tabular}[c]{@{}c@{}}Throughput\\(frames/s)\end{tabular}                                                         & 62.5               & 35.7               & {\bf --}       &   {\bf --}       & 69.9         & 46.7              & 96.2              & 63.5       & 136.4            \\ \hline
			GOPS                                                             & 1.19               & 0.94               & 0.48      &  0.74      & 1.33*        & 1.23*             & 2.15              & 1.67       & 1.33            \\ \hline
			\begin{tabular}[c]{@{}l@{}}Energy\\(mJ/frame)\end{tabular}            & \textgreater200*   & \textgreater500*   & {\bf --}   &  {\bf --}           & 150*         & 63                & 22.8                & 33.7      & 14.4               \\ \hline
		\end{tabular}
	}
	%\vspace{-1em}
\end{table}

{\bf Comparison with State-of-the-art:} Table~\ref{tab:energy_comp} compares {\em Synergy} with the recent FPGA-based CNN works. Note that CaffePresso~\cite{caffepresso_cnn} is using a development platform with significantly more resources, and is running at a higher clock speed. Moreover, as Darknet doesn't support data quantization feature and fixed-point implementation, {\em Synergy} uses 32-bit floating-point design that consumes much more resources than 32/16-bit fixed-point designs on FPGAs. As shown in Table~\ref{tab:energy_comp}, even though we have handicapped ourselves with floating-point operations, our implementations (both CIFAR\_full and MNIST) are superior to ~\cite{caffepresso_cnn}\cite{DeepBurning} in terms of throughput (frames per second), giga-operations-per-second (GOPS), and energy consumption. Compared to \cite{fpgaConvNet}, GOPS of our MNIST and MPCNN designs achieve 4.5x and 1.8x speedup, respectively. Table~\ref{tab:energy_comp} demonstrates that {\em Synergy} can provide high-throughput and energy-efficient mapping of CNN models on embedded heterogeneous SoC platforms.

%{\em Synergy} is somewhat limited by the lack of data quantization feature in {\em Darknet} to support fixed-point. The latency of the 32-bit floating-point implementation in {\em Synergy} is worse than fixed-point designs in Table~\ref{tab:energy_comp}. Moreover, {\em Synergy} currently cannot support large networks on embedded FPGAs due to the lack of data compression and quantization features in {\em Darknet}. We are working on extending {\em Darknet} with data compression and quantization support.

%In the experiment, we use this configuration with 6 HW accelerators in a big, fast cluster and the 2 HW accelerators in a smaller, slower cluster. Since the slower cluster is significantly slower than the faster cluster, we decide to use the NEON accelerators on the same job queue as that of the smaller cluster to help with the workload.
%In the experiment, we group the 6 faster HW accelerators into a big cluster, while the rest two go to a small cluster. As the big cluster is much faster than the small one, we decide to group the NEON accelerators into the small cluster.

%In the experiment, we use this configuration with 6 HW accelerators in a big, fast cluster and the 2 HW accelerators in a smaller, slower cluster. Since the slower cluster is significantly slower than the faster cluster, we decide to group the NEON accelerators  the smaller cluster to help with the workload.
%\todoinline{Add latency of single MM implementation}

\subsection{Advantage of Heterogeneity}
\label{subsec:impact_hetero}
\begin{sloppypar}

\begin{figure}[ht]
	\centering
	%\vspace{-0.5em}
	\includegraphics[width=0.6\columnwidth]{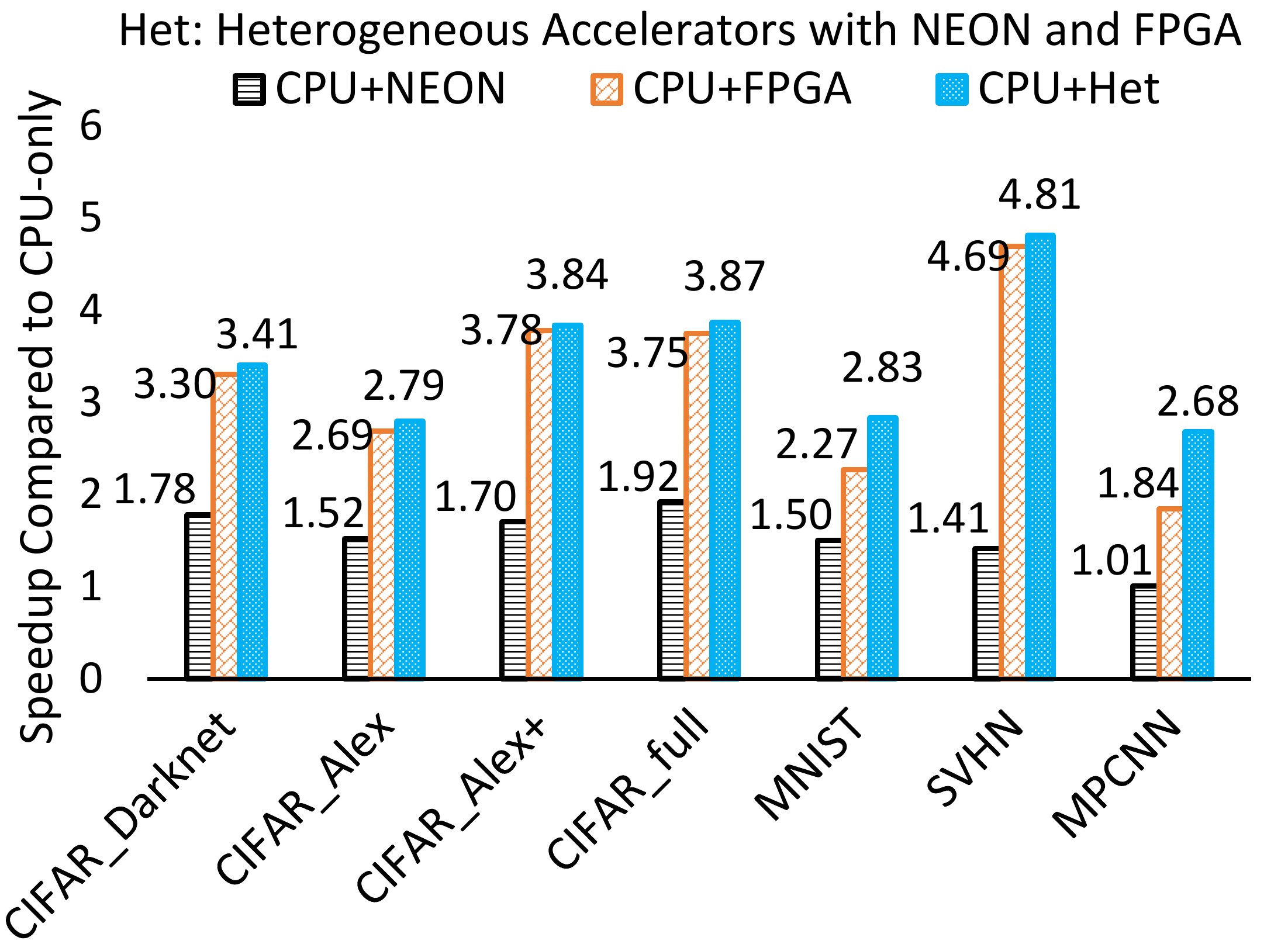}
	\caption{Latency Improvement with Accelerators Compared to CPU-only Solutions for Non-Pipelined Designs}
	%\vspace{-1em}
	\label{fig:lat_improvement}
\end{figure}

We now investigate the impact of heterogeneity in improving the CNN performance in {\em Synergy}. Figure~\ref{fig:lat_improvement} shows the latency of different non-pipelined CNN implementations (single-threaded, leveraging single-core ARM): {\em CPU+NEON}, {\em CPU+FPGA}, and {\em CPU+Het}, which consists of FPGA and NEON accelerators compared to the baseline single-core ARM design. Compared to the {\it CPU+FPGA} design, the heterogeneous implementation with FPGA and NEON {\it CPU+Het} improves the latency by 12\% on an average with 45\% maximum improvement for {\em MPCNN} model.

\begin{figure}[ht]
	\centering
	%\vspace{-0.5em}
	\includegraphics[width=0.6\columnwidth]{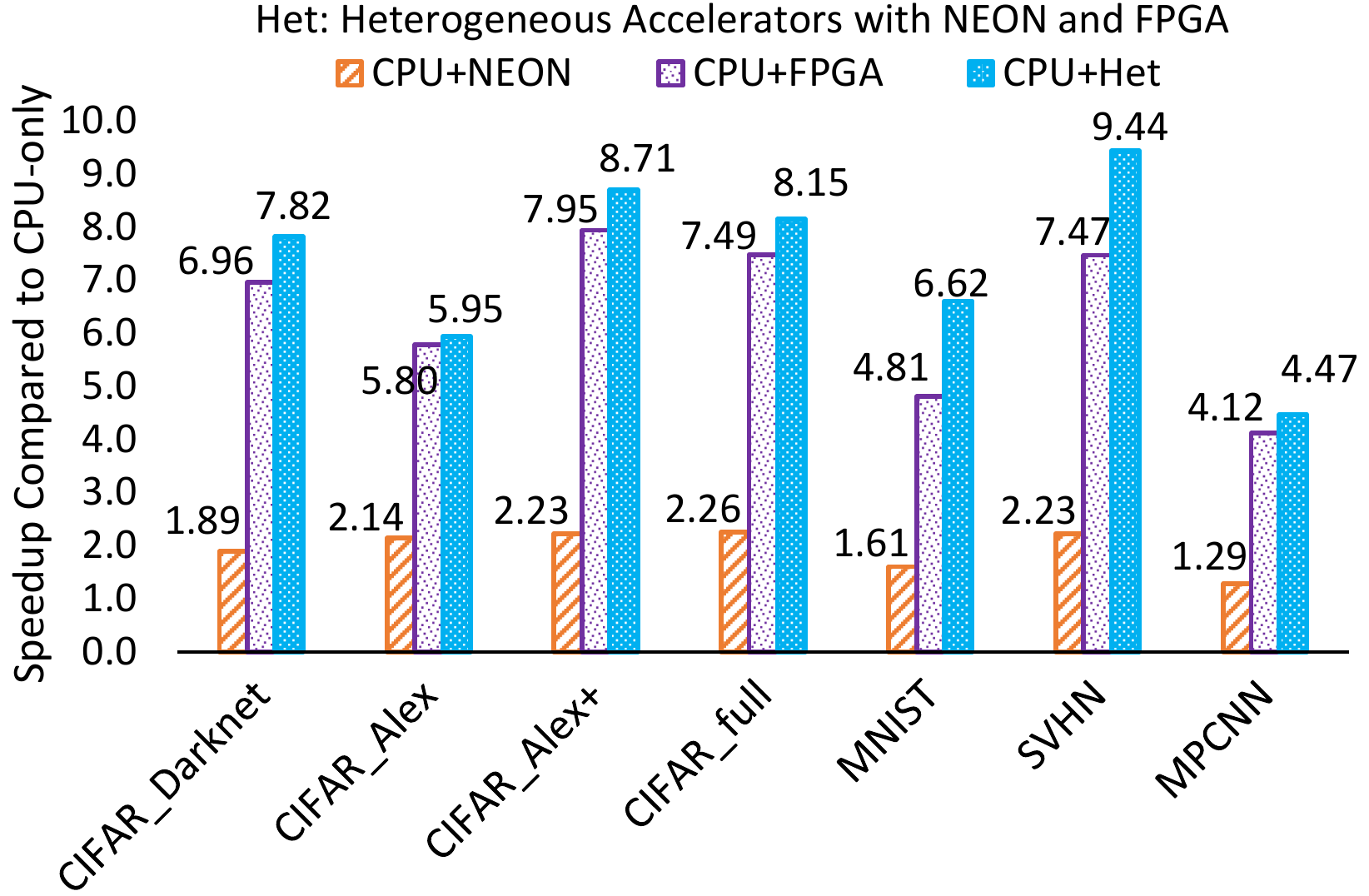}
	\caption{Throughput Improvement with Accelerators Compared to CPU-only Solutions for Pipelined Designs}
	%\vspace{-1em}
	\label{fig:throughput_improvement}
\end{figure}

The throughput speedup of different pipelined CNN implementations (multi-threaded, using two ARM cores): {\em CPU+NEON}, {\em CPU+FPGA}, and {\em CPU+Het} compared to the baseline single-core ARM design is shown in Figure~\ref{fig:throughput_improvement}. Compared to the {\it CPU+FPGA} designs, the heterogeneous implementations with FPGA and NEON {\it CPU+Het} achieves 15\% better throughput on an average (37\% maximum improvement for {\em MNIST} benchmark).

\end{sloppypar}

\subsection{Transparent Accelerators: Work Stealing}
\begin{figure}[ht]
	\centering
	%\vspace{-0.5em}
	\includegraphics[width=0.7\columnwidth]{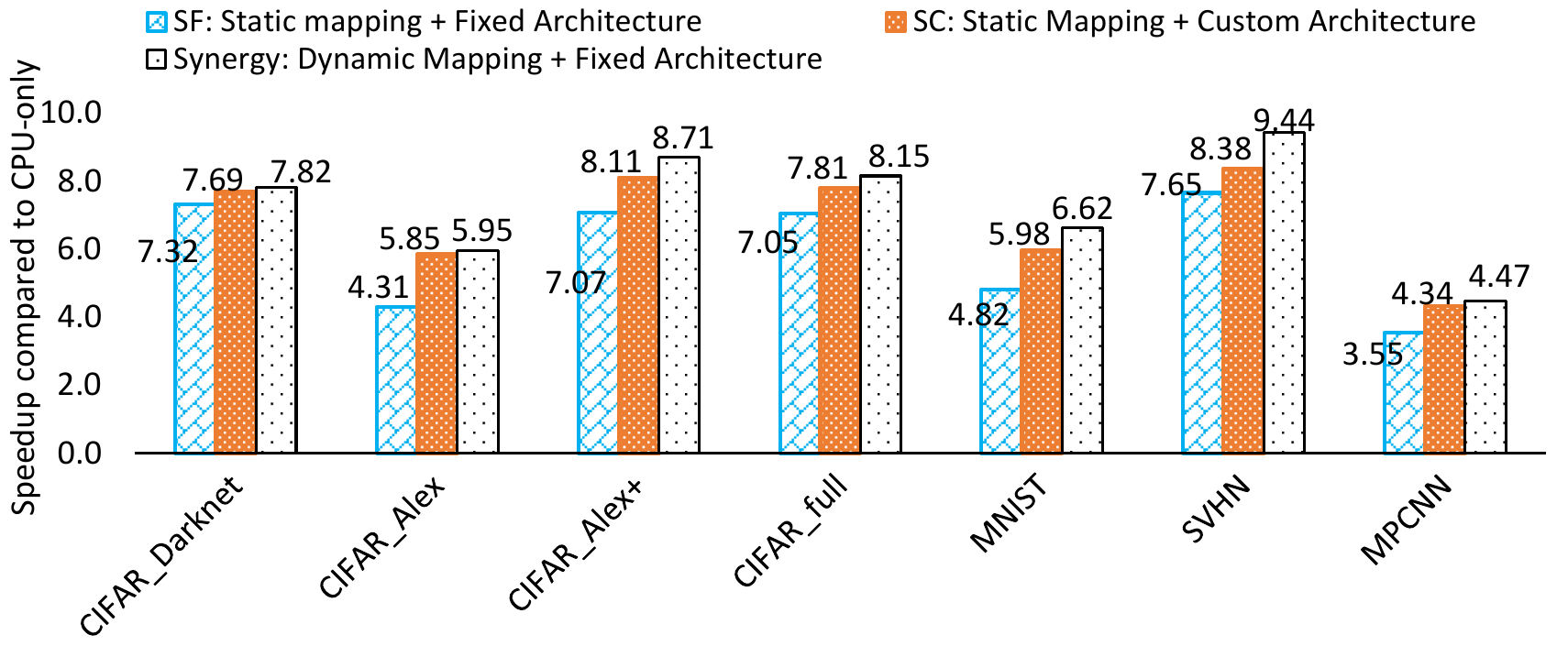}
	\caption{Advantage of Work stealing}
	%\vspace{-1em}
	\label{fig:transparent}
\end{figure}

\begin{sloppypar}
We show the advantage of dynamic load balancing across accelerators using work-stealing in {\em Synergy} versus static mapping of the CONV layers to the accelerators. We consider two different clusters and PE configurations for static mapping. The first cluster configuration consists of two clusters (Cluster-0: 2 NEONs + 2 S-PE; Cluster-1: 6 F-PE) used in {\em Synergy} across all benchmarks. But unlike {\em Synergy}, the CONV layers are statically assigned to the clusters based on their workload. We refer to this as {\it static-mapping+fixed-architecture} ({\it SF}). Figure~\ref{fig:transparent} shows that the  {\it SF} designs can achieve 6.1x better throughput compared to the well-optimized CPU designs. 

%We show the advantage of dynamic load balancing across accelerators using work-stealing in {\em Synergy} versus static mapping of the CONV layers to the accelerators. We consider two different clusters and PE configurations for static mapping. The first cluster configuration consists of two clusters (Cluster-0: 2 NEONs + 2 S-PE; Cluster-1: 6 F-PE) used in {\em Synergy} across all benchmarks. But unlike {\em Synergy}, the CONV layers are statically assigned to the clusters based on their workload. We refer to this as {\it static-mapping+fixed-architecture} ({\it SF}). Figure~\ref{fig:transparent} shows that the  {\it SF} designs can achieve 6.1x better throughput compared to the well-optimized CPU designs. Compared to the non-pipelined implementations, Table~\ref{tab:fpga_utilization} shows that the {\it SF} designs increase the accelerator cluster utilization to 92.7\% on average from 56.1\%. 

\begin{figure}[h]
	\centering
	%\vspace{-0.5em}
	\begin{subfigure}[b]{0.47\columnwidth}
		{
			\centering
			\includegraphics[width=0.7\columnwidth]{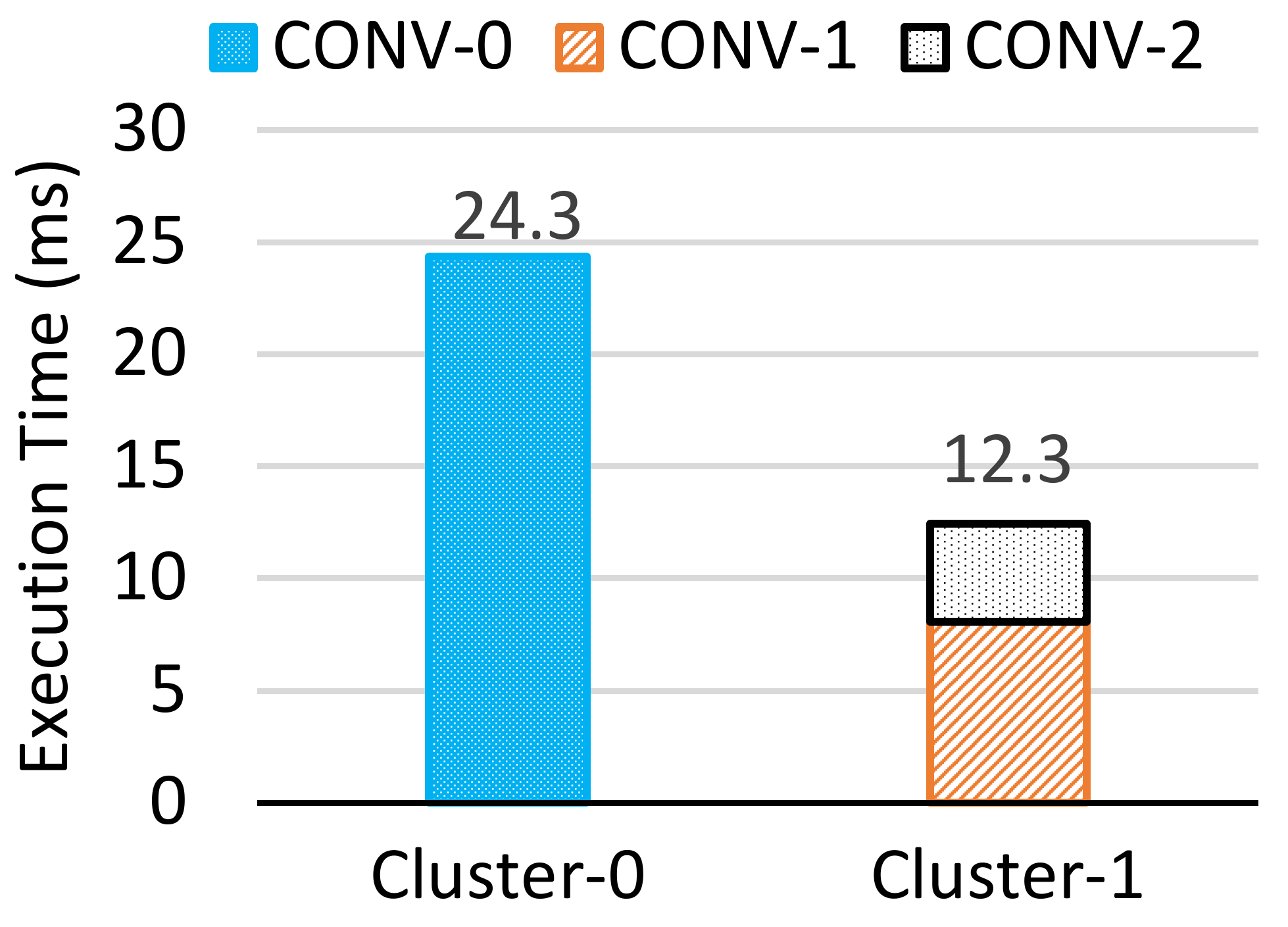}
			%\vspace{-1em}
			\caption{SF Configuration with Two Clusters}
			\label{fig:layer_time_imbalanced_cluster}
		}
	\end{subfigure}
	\quad
	\begin{subfigure}[b]{0.47\columnwidth}
		{
			\centering
			\includegraphics[width=0.7\columnwidth]{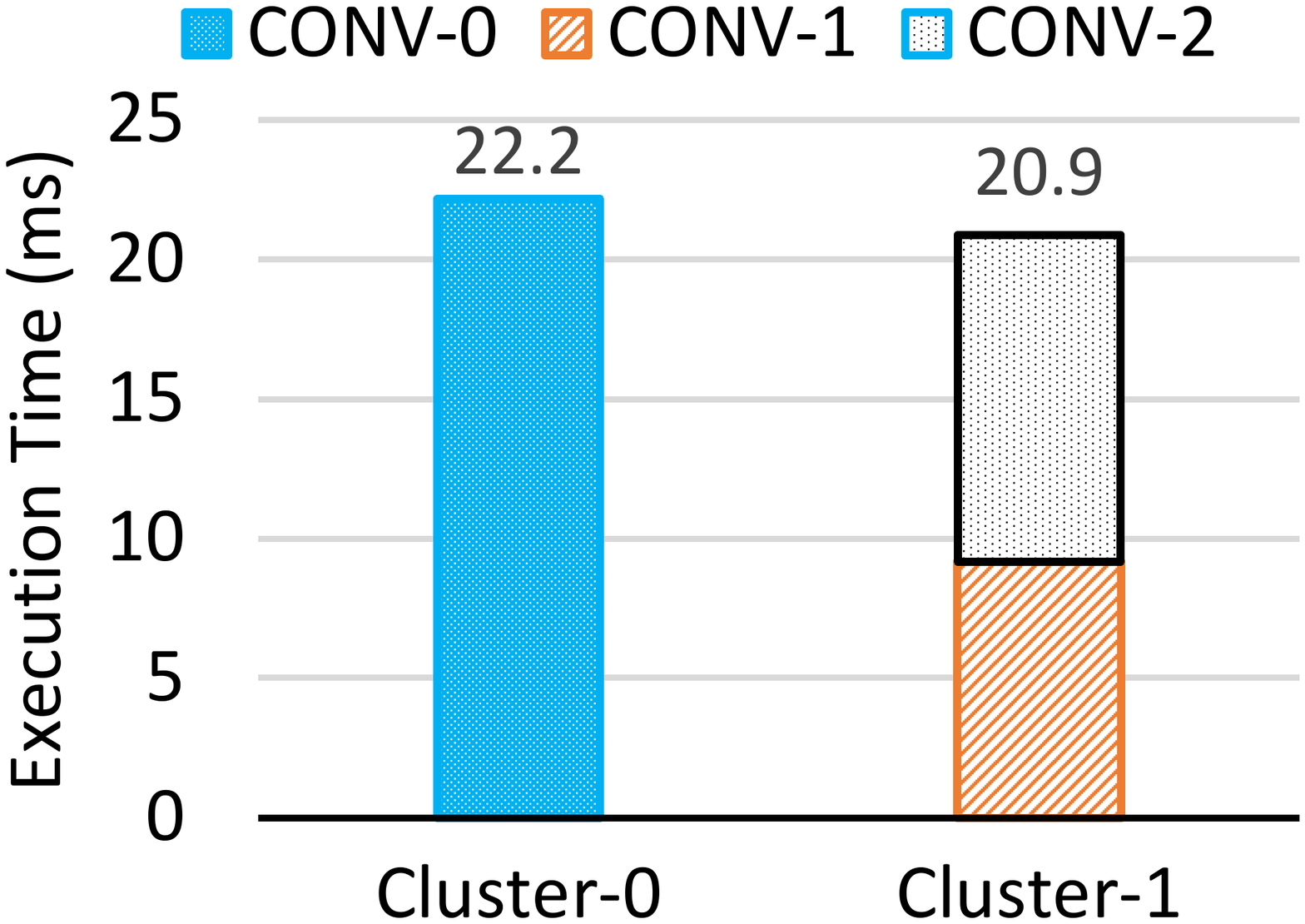}
			%\vspace{-1em}
			\caption{Synergy: same SF configuration + work-stealing}
			\label{fig:layer_time_balanced_cluster}
		}
	\end{subfigure}
	%\vspace{0.01em}
	\caption{Dynamic Load Balancing in CIFAR\_Alex. SF Conf.: Cluster-0 (2 NEONs + 2 S-PE), Cluster-1 (6 F-PE)}
	%\vspace{-1em}
	\label{fig:badMC_vs_goodMC}
\end{figure}

However, the {\it SF} designs are inefficient as the workload assigned to the clusters might not be balanced due to the different computation requirement of each CONV layer. Figure~\ref{fig:layer_time_imbalanced_cluster} presents the execution time of each CONV layer in {\em CIFAR\_Alex} model with this configuration. The CONV-0 layer is mapped to Cluster-0, while CONV-1 and CONV-2 layers are mapped to Cluster-1. As shown in Figure~\ref{fig:layer_time_imbalanced_cluster}, the runtime of Cluster-0 and Cluster-1 are 24.3 ms and 12.3 ms per frame, respectively. This imbalance in execution time between the clusters leads to poor cluster utilization and throughput.

{\em Synergy} employs work-stealing to automatically balance workload of different clusters. This provides a network-agnostic feature in {\em Synergy}, as the jobs from different CONV layers are automatically distributed across the different clusters to achieve self-balancing. With the same generic cluster configuration used in the {\it SF} designs, Figure~\ref{fig:transparent} shows that {\it Synergy} improves the throughput by average 24\% compared to the {\it SF} designs. The performance improvement comes from the balanced clusters. Figure~\ref{fig:layer_time_balanced_cluster} presents the execution time of each CONV layer of the {\it Synergy} design for {\em CIFAR\_Alex} benchmark. The runtime of Cluster-0 and Cluster-1 are 22.2 ms and 20.9 ms per frame, respectively. Compared to the {\it SF} design in Figure~\ref{fig:layer_time_imbalanced_cluster}, the workload of Cluster-0 and Cluster-1 are balanced.

%{\em Synergy} employs work-stealing to automatically balance workload of different clusters. This provides a network-agnostic feature in {\em Synergy}, as the jobs from different CONV layers are automatically distributed across the different clusters to achieve self-balancing. With the same generic cluster configuration used in the {\it SF} designs, Figure~\ref{fig:transparent} shows that {\it Synergy} improves the throughput by average 24\% compared to the {\it SF} designs. The performance improvement comes from the balanced clusters. Figure~\ref{fig:layer_time_balanced_cluster} presents the execution time of each CONV layer of the {\it Synergy} design for {\em CIFAR\_Alex} benchmark. The runtime of Cluster-0 and Cluster-1 are 22.2 ms and 20.9 ms per frame, respectively. Compared to the {\it SF} design in Figure~\ref{fig:layer_time_imbalanced_cluster}, the workload of Cluster-0 and Cluster-1 are balanced. Also {\it Synergy} achieve 99.8\% average cluster utilization as shown in Table~\ref{tab:fpga_utilization}.

\begin{table}[h]
	\centering
	%\vspace{-0.5em}
	\caption{Best Cluster Configurations for CNN Models under Static Mapping + Custom Architectures}
	\label{tab:best_cluster_config}
	%\resizebox{0.7\columnwidth}{!}{
	\begin{tabular}{|l|c|c|c|c|}
		\hline
		\multirow{2}{*}{Benchmarks} & \multicolumn{2}{c|}{Cluster 0} & \multicolumn{2}{c|}{Cluster 1} \\ \cline{2-5}
		& NEON     & FPGA IP             & NEON         & FPGA IP         \\ \hline
		CIFAR\_Darknet             & 0        & 2 S-PE + 1 F-PE     & 2            & 5 F-PE          \\ \hline
		CIFAR\_Alex                & 0        & 2 S-PE + 2 F-PE     & 2            & 4 F-PE          \\ \hline
		CIFAR\_Alex+               & 2        & 2 S-PE + 2 F-PE     & 0            & 4 F-PE          \\ \hline
		CIFAR\_full                & 0        & 2 S-PE + 2 F-PE     & 2            & 4 F-PE          \\ \hline
		MNIST                      & 2        & 2 S-PE + 2 F-PE     & 0            & 4 F-PE          \\ \hline
		SVHN                       & 2        & 2 S-PE + 2 F-PE     & 0            & 4 F-PE          \\ \hline
		MPCNN                      & 0        & 2 S-PE + 2 F-PE     & 2            & 4 F-PE          \\ \hline
	\end{tabular}
	%}
	%\vspace{-0.8em}
\end{table}

Finally, we show that {\em Synergy} work-stealing with generic cluster architecture is competitive and even better than CNN-model specific customized cluster configurations. We call this {\it static-mapping+custom-architecture} ({\it SC}) designs. In the {\it SC} designs, we find the best multi-cluster configuration for each CNN model by exploring all possible cluster configurations. The best multi-cluster configurations are shown in Table~\ref{tab:best_cluster_config}\footnote{The number of clusters in this work can be $t$, where $t \in \mathbb{N}$.}. The CONV layers are statically mapped to these clusters. Note that unlike optimized cluster configurations in {\em SC} designs, {\it Synergy} leverages the same generic cluster configuration used in the {\it SF} designs for various CNN models.  As shown in Figure~\ref{fig:transparent}, {\it Synergy} still achieves 6\% better throughput than {\it SC} designs.  This is because in the static mapping approaches ({\it SF} and {\it SC}) an entire CONV layer is assigned to a cluster and it is hard to perfectly balance the cluster workloads. In contrast, the work-stealing in {\em Synergy} at the granularity of job-level (tiled MM) can easily balance the workload even with un-optimized generic accelerators. The {\it work stealing} feature in {\em Synergy} empowers developers to easily switch between different networks at runtime without losing performance.

\begin{table}[t]
	\centering
	\caption{Accelerator Cluster Utilization Comparison Across SF, SC and Synergy}
	\label{tab:fpga_utilization}
	%\resizebox{0.8\columnwidth}{!}{
	% \usepackage{multirow}
	\begin{tabular}{|c|c|c|c|c|}
		\hline
		\multirow{2}{*}{Benchmarks} & \multirow{2}{*}{\begin{tabular}[c]{@{}c@{}}Non-\\pipelined (\%)\end{tabular}} & \multicolumn{3}{c|}{Pipelined (\%)} \\ \cline{3-5}
		&                                    & SF         & SC        & Synergy        \\ \hline
		CIFAR Darknet               & 50.77                              & 95.32      & 97.55     & 99.89     \\ \hline
		CIFAR Alex                  & 53.56                              & 92.72      & 96.61     & 99.83     \\ \hline
		CIFAR Alex+                 & 61.28                              & 98.81      & 98.73     & 99.95     \\ \hline
		CIFAR full                  & 54.06                              & 93.53      & 94.97     & 100.00     \\ \hline
		MNIST                       & 59.03                              & 85.63      & 96.09     & 99.89     \\ \hline
		SVHN                        & 53.00                              & 94.72      & 96.86     & 99.26     \\ \hline
		MPCNN                       & 60.62                              & 86.47      & 94.45     & 99.79     \\ \hline
		mean                        & 56.05                              & 92.46      & 96.47     & 99.80      \\ \hline
	\end{tabular}
	%}
	%\vspace{-1em}
\end{table}

To better understand the performance improvement, Table~\ref{tab:fpga_utilization} shows the accelerator cluster utilization of various designs. The non-pipelined designs are the best single-threaded implementations (the blue bars in Figure~\ref{fig:lat_improvement}) leveraging single-CPU, NEON core and FPGA accelerators. As shown in Table~\ref{tab:fpga_utilization}, the cluster utilization of the non-pipelined designs is very low, indicating FPGA being idle for 43.95\% (=$1-56.05\%$) of the total execution time on average. The reason is that in non-pipelined design, FPGA accelerators have to wait for CPU or NEON core to finish their work. With multi-threading support, the pipelined designs significantly increase the cluster utilization (above 90\%), as various computing elements can work simultaneously. Table~\ref{tab:fpga_utilization} shows that the {\it SF} designs increase the accelerator cluster utilization to 92.5\% on average from 56.1\%. Compared to the {\it SF} designs, the cluster utilization of the {\it SC} designs achieves 96.5\% averaged across the benchmarks. This is because the {\it SC} designs use the fine-tuned cluster configurations and workload assigned to the clusters is more balanced. As mentioned above, since {\em Synergy} leverages the work-stealing scheduler which works at the finer granularity of job-level (tiled MM), the scheduler helps to improve the cluster utilization at runtime by balancing workload in clusters. The average cluster utilization of {\it Synergy} achieves 99.8\% as shown in Table~\ref{tab:fpga_utilization}.

\end{sloppypar}

%% file: conclusion.tex
\section{conclusion}
%This paper presents Synergy, an automated hardware/software multi-threaded CNN inference framework on Xilinx Zynq SoC to fully utilize its heterogeneity. The proposed framework considers both intra-frame (multiple hardware accelerators and NEON cores on convolutional layers) and inter-frame (cooperative hardware/software threads on all layers) parallelism in CNN models. Moreover, we develop a work-stealing scheduler to improve load balance among convolutional layers and further increase throughput of CNN models. Our result shows that Synergy achieves 7.4x speedup, averaged across five representative CNN models, over an well-optimized software-only solution. Compared to the contemporary CNN implementations on the same SoC platform, Synergy delivers better throughput as well as energy-efficiency.

This paper presents Synergy, an automated, transparent hardware-software co-designed CNN inference framework on an embedded FPGA-based heterogeneous SoC architecture. Synergy fully utilizes the heterogeneity by leveraging diverse computing resources (CPUs, NEONs and FPGA) to accelerate CNNs. Moreover, Synergy provides a {\it work-stealing} scheduler in software to automatically balance the workload of accelerators, so that it can easily adapt to various networks at runtime without changing hardware or software implementations. Our result shows that Synergy achieves 7.3x speedup, averaged across seven representative CNN models, over a well-optimized software-only solution. Compared to the contemporary CNN implementations on the same SoC platform, Synergy delivers better throughput as well as energy-efficiency.
%\vspace{-1.0em}